\documentclass[review]{elsarticle}

\usepackage{geometry}
\usepackage[T1]{fontenc}

\usepackage{color}
\usepackage[normalem]{ulem}

\usepackage{xpatch}
\usepackage{amsmath}
\xpatchcmd{\MaketitleBox}{\hrule}{}{}{}
\xpatchcmd{\MaketitleBox}{\hrule}{}{}{}

\usepackage{lineno,hyperref}
\modulolinenumbers[5]

\usepackage{cleveref}
\usepackage{braket}

\newlength{\myl}%
\newcommand{\ZT}[1]{\textquotedblleft#1\textquotedblright}%
\newcommand{\rt}{(\vec{r},t)}%
\newcommand{\Nabla}{\vec{\nabla}}%
\newcommand{\dif}{\mathrm{d}}%
\newcommand{\INT}[3]{\settowidth{\myl}{$\displaystyle\int_{#1}^{#2}$}{\int_{#1}^{#2}\;\;\;\hspace{-\the\myl}\dif #3}\,}%

\biboptions{authoryear}

\begin{document}
\begin{frontmatter}
\title{The Five Problems of Irreversibility}
\author{Michael te Vrugt$^{1,2,3}$\\[1ex]
\textit{\small{$^1$ Institut f\"ur Theoretische Physik, $^2$Center for Soft Nanoscience, $^3$Philosophisches Seminar,}}\\\textit{\small{ Westf\"alische Wilhelms-Universit\"at M\"unster, D-48149 M\"unster, Germany}}\\[1ex]
This is the accepted manuscript of an article that has been published in a peer-reviewed journal.\\ When citing it, please always refer to the published version:\\[1ex]
Michael te Vrugt, \textit{The five problems of irreversibility}, Studies in History and Philosophy of Science \textbf{87}, pp. 136-146 (2021), https://doi.org/10.1016/j.shpsa.2021.04.006}
\date{}

\begin{abstract}
Thermodynamics has a clear arrow of time, characterized by the irreversible approach to equilibrium. This stands in contrast to the laws of microscopic theories, which are invariant under time-reversal. Foundational discussions of this \ZT{problem of irreversibility} often focus on historical considerations, and do therefore not take results of modern physical research on this topic into account. In this article, I will close this gap by studying the implications of \textit{dynamical density functional theory} (DDFT), a central method of modern nonequilibrium statistical mechanics not previously considered in philosophy of physics, for this debate. For this purpose, the philosophical discussion of irreversibility is structured into five problems, concerned with the source of irreversibility in thermodynamics, the definition of equilibrium and entropy, the justification of coarse-graining, the approach to equilibrium and the arrow of time. For each of these problems, it is shown that DDFT provides novel insights that are of importance for both physicists and philosophers of physics.
\end{abstract}


\end{frontmatter}

\tableofcontents
\newpage

\section{Introduction}
The temporal asymmetry of thermodynamics is one of the central problems in philosophy of physics. If a cup of hot coffee stands in a room, it will cool down until it has room temperature, but it will not spontaneously heat up by extracting heat from its environment. This is commonly seen as a consequence of the second law of thermodynamics, which assigns to each of these systems a quantity known as \ZT{entropy} that increases in these processes and that, most importantly, cannot decrease. Often, this is considered one of the most fundamental laws of physics \cite[p. 540]{Callender2001}.

However, as is also well-known, there is a conflict with the microscopic laws governing the motion of the individual particles that a macroscopic system consists of. These laws are invariant under time-reversal\footnote{When it comes to the most fundamental laws, this is not strictly true. The standard model of particle physics is invariant under CPT, which is a combination of charge-conjugation, mirror reflection and time-reversal. However, this effect is too small to account for the temporal asymmetry of thermodynamics \cite[p. 270]{Wallace2013}.}, which means that if a process can occur in one direction of time, it can also occur in the other direction. Thus, a cup of coffee at room temperature that spontaneously heats up would be in perfect agreement with the microscopic laws of physics, which makes it very difficult to explain why such a behaviour is never observed.

An intense discussion on this problem has emerged in philosophy of physics. It has evolved into a variety of sub-debates concerned with different explananda that, in this article will be classified into five problems. Often, foundational discussions of statistical mechanics focus on historical aspects. Moreover, there has been a growth of interest in formal approaches to coarse-graining based on the projection operator formalism \citep{Zwanzig1960,Mori1965}. There is, however, a lack of work that considers the implications of modern research on nonequilibrium statistical mechanics for foundational problems \citep{Wallace2015}. 

In this work, I will close this gap by providing the first philosophical discussion of \textit{dynamical density functional theory} (DDFT) \citep{Evans1979,MarconiT1999,teVrugtLW2020}, which is one of the cornerstones of modern statistical physics. DDFT has originally been developed for modeling simple and complex fluids. It is now a central method of theoretical soft matter physics and has, moreover, found applications in other fields such as biology \citep{AngiolettiBD2018}, chemistry \citep{LiuL2020}, epidemiology \citep{teVrugtBW2020}, or plasma physics \citep{DiawM2015}. Moreover, it is intimately connected to the projection operator formalism \citep{EspanolL2009,WittkowskiLB2012}, such that it provides a natural link between foundational debates in philosophy of physics and practical work in condensed matter physics. For each of the five sub-problems discussed in this work, it is shown that DDFT can make interesting contributions to the debate in philosophy.

This article is structured as follows: In \cref{theproblem}, I explain how \ZT{the} problem of irreversibility arises, and structure it into five different sub-problems. An introduction to DDFT can be found in \cref{ddft}. In \cref{one,two,three,five,four}, I explain in detail each problem and the contribution that DDFT can make to its solution. I conclude in \cref{conclusion}.

\section{\label{theproblem}The Five Problems of Irreversibility}
In this section, I will briefly recapitulate what is commonly referred to as \ZT{the problem of irreversibility}. This is a longwithstanding debate in the foundations of statistical mechanics, which has evolved into a variety of sub-debates that are dealing with different aspects. A careful distinction of the different explananda is required both to ensure a conceptually precise treatment and to appreciate that studying DDFT is important for a variety of foundational debates. For this purpose, I will introduce here a scheme of \ZT{problems of irreversibility}, which is used to provide structure both for the overview over the general debate and for the discussion of DDFT in the subsequent chapters. These will also present the five problems in more detail. Given that DDFT is a classical theory (although it can be connected to quantum-mechanical methods, see \citet{teVrugtLW2020} for a discussion), I will restrict myself to classical statistical mechanics here.

The first theory that is relevant is thermodynamics, which is a phenomenological theory that describes macroscopic systems\footnote{By a \textit{macroscopic system}, I mean a system that contains about $10^{23}$ particles. Strictly speaking, the thermodynamic limit requires that a macroscopic systems has infinitely many particles \citep{ThieleFHEKA2019}. This idealization, however, leads to additional philosophical problems \citep{MenonC2013} that are not relevant here.} A central observation in thermodynamics is that isolated systems tend to approach a state of \textit{thermodynamic equilibrium}, in which its macroscopic properties are approximately constant. This will, in practice, be the case only on certain observational timescales \cite[p. 545]{Callender2001}. For example, a cup of hot coffee that stands in a cold room will cool down until it has room temperature (equilibrium state).

This is associated with a quantity known as \textit{entropy}, which according to (a common interpretation of) the \textit{second law of thermodynamics} can never decrease in an isolated system. For example, it is not possible that the coffee spontaneously heats up by absorbing heat from the cold room, since in this case entropy would decrease. The tendency to approach equilibrium is often associated with the second law of thermodynamics, although \citet{BrownU2001} have argued that it has a more fundamental status and should be viewed as a \textit{minus first law}. This leads us to the first problem (\cref{one}), which is concerned with how the observation that macroscopic systems exhibit irreversible behavior is built into the axioms of thermodynamics.

The second theory we require is classical mechanics, which describes the microscopic dynamics of the individual particles. These are described by \textit{Hamilton's equations}, which are the fundamental laws of classical mechanics and have the important property of being \textit{time-reversal invariant}: If we record a movie of a process allowed by classical mechanics and play the movie backwards, then what we see will also be a process allowed by classical mechanics. The microscopic laws know no preferred direction of time. The microstate of a system is a point in the so-called \textit{phase space}, which is the central playground of classical mechanics.

Thus, there seems to be a contradiction between the microscopic laws of classical mechanics and the macroscopic laws of thermodynamics. In thermodynamics, systems are expected to evolve towards an equilibrium state, accompanied by an increase of entropy. This process is \textit{irreversible}, since its time-reversal would involve a decrease of entropy and is therefore forbidden by the second law. The laws governing the microscopic constituents of the system, however, are time-reversal invariant and would therefore allow for such a process. 

The connection between the behaviour of the individual particles and the dynamics of the macroscopic system it consists of is studied in \textit{statistical mechanics}. It can be constructed in two frameworks, which are referred to as the \textit{Boltzmann approach} and the \textit{Gibbs approach}. I present them here following \citet{Frigg2008}. In the Boltzmann approach, a system is, on the macroscopic level, characterized by a small number of variables (such as temperature and volume), which define the \textit{macrostate} of the system. Its \textit{microstate}, on the other hand, is characterized by the phase-space coordinates of all particles the system consists of. The microstate uniquely determines the macrostate, while the converse is not true. The \textit{entropy} of a macrostate is then introduced as a measure for the volume of the phase-space region corresponding to this macrostate. One can then introduce the equilibrium state as the state with maximal Boltzmann entropy. 

In the framework of Gibbs, one studies many-particle systems using \textit{ensembles}. These are hypothetical sets of infinitely many copies of the system which evolve according to the same laws, but with different initial conditions. One then introduces a probability density that for each point in phase space, i.e., for each microscopic state, gives the probability that a system that is randomly chosen from the ensemble is in this state (see \citet{Frigg2008} and \citet[p. 107]{FriggW2020}). It is helpful to think of this probability density as a fluid. In this framework, the entropy can then be given a microscopic definition as a function of the density - intuitively, it measures the volume of the fluid. The problem is now that one can easily prove using Hamilton's equations that the volume of this \ZT{fluid} is constant, a result known as \textit{Liouville's theorem}. This implies, of course, that the entropy is also constant - in conflict with thermodynamics, which demands that it increases during the approach to equilibrium. In the equilibrium state, one has a stationary phase-space distribution that maximizes the Gibbs entropy.

In physics, this problem has not gone unnoticed, and is routinely solved through a procedure known as \textit{coarse-graining}, which has originally been suggested by \citet{Gibbs1902} (see also \citet[pp. 550 - 551]{Robertson2018}). The starting point - in usual treatments - is that we are trying to describe macroscopic observations. Such observations do not allow to distinguish between certain microstates - if we shift a certain fluid molecule by a few nanometers, this will change the microstate of a bucket of full of water, but it will not make an observable difference. Therefore, we can simply group together the macroscopically indistinguishable microstates and average over them. Notably, this was done with an epistemic - and not ontological - justification. For the coarse-grained distribution, Liouville's theorem does not hold. Hence, we can define, from the coarse-grained distribution, a coarse-grained entropy which measures its volume. This entropy is then allowed to increase. The coarse-graining can be mathematically implemented in various ways. I will denote by \textit{coarse-graining} any procedure that involves replacing the microscopic (\textit{fine-grained}) distribution function of the system by an averaged one.

Procedures of this form are routinely and successfully used in physics to derive irreversible macroscopic transport equations from the underlying Hamiltonian dynamics. Nevertheless, they are discussed very controversially in philosophy. The objection is (roughly) that we have, through the coarse-graining, artificially introduced an asymmetry that has not been there originally. This has been justified by an appeal to our limited capability of observation, but the increase of entropy is a physical effect that should not depend on how good our microscopes are \cite[pp. 563 - 565]{Robertson2018}. In fact, this discussion is - as we will see below - concerned with two different problems. On the one hand, we can ask whether the fine-grained or the coarse-grained definition of entropy and equilibrium is the \ZT{correct} one (\cref{two}), on the other hand, we can ask how the coarse-graining procedure can be justified (\cref{three}).

Finally, we need to find a way to derive - with or without coarse-graining - the irreversible macroscopic dynamics from the reversible microscopic dynamics. In statistical mechanics, there are well-established procedures for this purpose, which are often based on the argument that, for a nonequilibrium initial condition, it is far more likely to move to equilibrium than away from it, since the equilibrium state is the one with the largest phase-space volume \cite[pp. 321-323]{North2011}. The problem is that such arguments can - because of the time-reversal invariance of the underlying laws - also be applied to the past \cite[p. 363]{Callender1999}. This is a problem because our records of the past tell us that entropy used to be lower. A solution that is very popular in philosophy is the \textit{past hypothesis}. The idea is that, since the asymmetry between past and future cannot be a consequence of the dynamical laws of the universe, it has to be a consequence of the boundary conditions. In particular, if we assume that the entropy of the early universe was very low, we could have an explanation for why it increases afterwards \citep{Callender2016}.

It is an interesting observation that, despite the fact that the entropy of the early universe plays such a prominent role in foundational discussions of statistical mechanics, it is typically of no importance in the everyday life of a statistical physicist who derives irreversible transport equations. The reason is that we are, once again, dealing with different questions. On the one hand, we can ask ourselves why systems, given an initial nonequilibrium state, tend to approach equilibrium irreversibly (\cref{four}). On the other hand, we should ask ourselves why the explanation we have found for this effect does not apply to the past (\cref{five}). Distinctions of this form have been expressed by Boltzmann (see \cite[p. 530]{BrownU2001}), \citet[p. 218]{Penrose1994} and \citet[pp. 47-48]{Price1996}.

In summary, the five problems are:

\begin{itemize}
    \item \textbf{What is the source of irreversibility in thermodynamics? (Q1)}
    
    This question asks for the source of irreversibility \textit{within thermodynamics}, i.e., purely on the level of the macroscopic theory. It is controversial which of the laws of thermodynamics is actually responsible here.
    
    \item \textbf{How should \ZT{equilibrium} and \ZT{entropy} be defined? (Q2)}
    
    Here, we look for a definition of the explanandum. The questions \ZT{Why do systems approach equilibrium?} and \ZT{Why does entropy increase?} are answered differently by persons who have different ideas on what equilibrium and entropy are, which is a common source of confusion.
    
    \item \textbf{(How) Can coarse-graining be justified? (Q3)}
    
    Since coarse-graining, which is frequently used in explanations of the approach to equilibrium, is a controversial procedure, it needs to be clarified whether and how it can be justified. This is a \textit{separate} problem, since coarse-graining is also used in situations without a relation to thermodynamic irreversibility.
    
    \item \textbf{Why do systems approach equilibrium? (Q4)}
    
    This question asks why systems, given an initial nonequilibrium state, (irreversibly) approach equilibrium.
    
    \item \textbf{Why does the arrow of time have the direction it has? (Q5)}
    
    Here, we look for an explanation for why our answer to Q4 does not apply to the past, i.e., we wish to find out why the entropy always increases from past to future.
\end{itemize}

A popular framework for discussing the microscopic origin of irreversible dynamics is the projection operator method. In this work, I will refer to this approach as \textit{Mori-Zwanzig formalism}, as is common in physics\footnote{Approaches based on projection operators are known by a variety of names. In addition to \ZT{Mori-Zwanzig formalism}, one also finds, e.g., \ZT{Zwanzig-Zeh-Wallace framework} \citep{Robertson2018} - referring to the conceptual discussions of this formalism by \citet{Zeh1989} and \citet{Wallace2015} -  \ZT{Kawasaki-Gunton operator method} \citep{Yoshimori2005} - referring to \citet{KawasakiG1973} - or \ZT{Mori-Zwanzig-Forster technique} \citep{WittkowskiLB2012,WittkowskiLB2013} - referring also to \citet{Forster1974}. While these names sometimes refer to slightly different forms, the general idea is always the same.} \citep{teVrugtW2019}. The general idea is to introduce a projection operator to project the full dynamics onto the part that only depends on the \ZT{relevant} degrees of freedom of the system. One then derives a formally exact transport equation that contains a term depending on the irrelevant degrees of freedom at the initial time $t_0$ and a term depending on the state of the system at previous times. To obtain a closed memoryless transport equation - that is irreversible! - it is then assumed that the irrelevant part of the density vanishes at $t_0$ and that memory effects can be ignored (Markovian approximation) \citep[pp. 553-556]{Robertson2018}. The original formalism was developed by \citet{Nakajima1958}, \citet{Zwanzig1960}, and \citet{Mori1965}, and has subsequently been extended to incorporate, for example, the dynamics of fluctuations \citep{Grabert1978} and time-dependent Hamiltonians \citep{teVrugtW2019,MeyerVS2019}. An introduction to this method can be found in \citet{teVrugtW2019b}, a general overview in \citet{Grabert1982}.

\section{\label{ddft}Dynamical density functional theory}
After having presented and analyzed the discussion of thermodynamic irreversibility in philosophy of physics, we can now turn to an example of an irreversible theory used in modern physics, namely \textit{dynamical density functional theory} (DDFT). Here, I give a brief overview. A detailed review can be found in \citet{teVrugtLW2020}.

DDFT is a theory for the nonequilibrium dynamics of classical many-body systems. It is an extension of classical \textit{density functional theory} (DFT), which is a highly successful and formally exact theory (based on the quantum-mechanical DFT by \citet{HohenbergK1964}) that allows to find the equilibrium state of a many-particle system. This is done by minimizing a free energy functional that depends on the one-body density $\rho(\vec{r})$ (which gives the probability of finding a particle at position $\vec{r}$). In DDFT (which is an approximate theory), this is extended to the out-of-equilibrium case by assuming that the nonequilibrium system is driven towards the state in which the free energy functional $F$ is minimized. This leads to the governing equation of deterministic DDFT, given by
\begin{equation}
\frac{\partial}{\partial t}\rho\rt=\Gamma\vec{\nabla}\cdot\bigg(\rho\rt\vec{\nabla}\frac{\delta F[\rho]}{\delta\rho\rt}\bigg)
\label{trddft}%
\end{equation}
with time $t$ and mobility $\Gamma$. Here, $\delta$ denotes a functional derivative and $\Nabla$ is the del operator. 

The DDFT equation \eqref{trddft} was originally proposed by \citet{Evans1979} based on phenomenological arguments, and later derived from microscopic dynamics by \citet{MarconiT1999,MarconiT2000} and \citet{ArcherE2004}. DDFT also exists in the stochastic form 
\begin{equation}
\frac{\partial}{\partial t}\rho\rt=\Gamma\Nabla\cdot\bigg(\rho\rt\Nabla\frac{\delta F[\rho]}{\delta\rho\rt}\bigg) + \Nabla\cdot\bigg(\sqrt{2\Gamma k_B T\rho\rt}\vec{\eta}\rt\bigg),
\label{trddfts}
\end{equation}
where $k_B$ is the Boltzmann constant, $T$ is the temperature, and $\vec{\eta}\rt$ is a multiplicative noise. Stochastic DDFT was, in various forms, developed by \citet{Munakata1989}, \citet{Fraaije1993}, \citet{Kawasaki1994} and \citet{Dean1996}. (The relation between stochastic and deterministic DDFT is discussed in \cref{two,three}.) Today, DDFT is among the most widely used methods in nonequilibrium statistical mechanics, with applications including biological swimmers \citep{MenzelSHL2016}, disease spreading \citep{teVrugtBW2020}, ions in capacitors \citep{BabelEL2018}, plasmas \citep{DiawM2016}, thin films \citep{RobbinsAT2011}, tumor growth \citep{AlSaediHAW2018}, and much more. Moreover, a variety of extensions have been developed that allow to model, e.g., flow fields \citep{RauscherDKP2007}, hydrodynamic interactions \citep{RexL2008}, nonspherical and active particles \citep{WittkowskiL2011}, nonisothermal systems \citep{WittkowskiLB2012}, or systems with strict particle order \citep{WittmannLB2020}.

Among the main fields of application of DDFT are colloidal and atomic fluids. Colloidal fluids consist of large particles (colloids) that are immersed in a fluid consisting of many small particles (solvent). Due to the presence of the solvent, the motion of the particles is overdamped (i.e., momentum degrees of freedom can be neglected), and subject to noise. The motion of the colloids is described by the Langevin equations
\begin{equation}
\frac{\dif \vec{r}_i(t)}{\dif t}=\Gamma\vec{F}_i(t) + \vec{\chi}_i(t),
\label{langevin}%
\end{equation}
where $\vec{r}_i$ is the position of the $i$-th particle, $\vec{F}_i$ the force acting on it, and $\vec{\chi}_i(t)$ an additive white noise term with zero mean. As shown by \citet{MarconiT1999}, the DDFT equation \eqref{trddft} can be derived from the Langevin equations \eqref{langevin} based on the assumption that the pair correlation of the nonequilibrium system is identical to that of an equilibrium system with the same one-body density (adiabatic approximation). Atomic fluids, on the other hand, consist of atoms, i.e., of only one type of particle. Their microscopic dynamics is given by Hamilton's equations, which are undamped, rather than by the overdamped Langevin equations. Deriving a DDFT for one-component fluids therefore requires an elimination of the momentum degrees of freedom, which is possible by assuming that the momentum relaxes much faster than the density \cite[p. 47]{Kawasaki1994}.

The Mori-Zwanzig formalism (projection operator method) is an important tool in DDFT, as it allows for a systematic derivation of the DDFT equations \eqref{trddft} and \eqref{trddfts} from the microscopic dynamics of the individual particles. (This is discussed in detail in \citet{teVrugtLW2020}.) Consequently, the projection operator method allows to relate the physical research on DDFT to the discussion of irreversibility in philosophy of physics. Microscopic derivations of DDFT from the reversible Hamiltonian dynamics were presented by \citet{Yoshimori2005} and \citet{EspanolL2009}. Moreover, one can obtain DDFT from a stochastic underlying dynamics through additional projections \citep{Kawasaki1994,EspanolV2002}. An important application of projection operators is the derivation of generalizations of DDFT that incorporate additional order parameter fields. For example, the energy density is used as a relevant variable in addition to the number density in \ZT{extended dynamical density functional theory} (EDDFT), which was derived using the projection operator formalism by \citet{WittkowskiLB2012,WittkowskiLB2013}.

\section{\label{one}First Problem: The Source of Irreversibility in Thermodynamics}
As discussed in \cref{theproblem}, entropy and irreversibility are studied in two fields of physics: Thermodynamics, which is a phenomenological theory of macroscopic states, and statistical mechanics, which describes macroscopic systems by developing a statistical description of their microscopic constituents. One of the aims of statistical mechanics - though by no means the only one - is to provide a microscopic justification for the principles that are introduced as axioms on the level of thermodynamics. Here, we are interested in a particular property of thermodynamics, namely the existence of irreversible processes. If we wish to figure out where in the transition from microscopic to macroscopic physics these come into play, a good way to start is therefore to first figure out in which of the axioms of thermodynamics irreversibility can be found, and then to have a look at the microscopic foundations of that particular axiom.

\citet[pp. 313-315]{Uffink2001} attributes the difficulty of locating the temporal asymmetry in thermodynamics to the absence of equations of motion. A typical understanding of time-reversal symmetry is that a theory has this symmetry if, when we time-reverse a temporal evolution that is allowed by the theory, the resulting (backwards) evolution would also be allowed by the theory. It is not immediately clear how to apply this definition to a theory that is not primarily concerned with time evolutions.

Thermodynamics is based on four axioms (\ZT{Haupts\"atze}), namely the zeroth law (transitivity of thermal equilibrium), the first law (conservation of energy), the second law (entropy cannot decrease) and the third law (a system cannot reach zero temperature). In physics textbooks, the second law typically gets credit for irreversibility. This law was stated in \cref{theproblem} as \ZT{Entropy cannot decrease in a closed system}. Although this is a very common understanding, it is very difficult to state the content of the second law in a precise way since it exists in many different forms that are not fully equivalent and have differing connections to the arrow of time. \citet[p. 306]{Uffink2001}, who has analyzed this point in detail, has compared the interpretation of the second law to the interpretation of a work from Shakespeare. 

The relation of the second law to irreversibility depends, as shown by \citet{Uffink2001} in a detailed historical analysis, on its precise formulation. In an influential article, \citet{BrownU2001} have argued that the temporal asymmetry of thermodynamics is more fundamental than and logically prior to the second law. Therefore, they propose a \ZT{minus first law} that should be added to the standard set of laws. It states that for each isolated system there exists a unique state of equilibrium that it will spontaneously enter. The time-asymmetry of \textit{thermodynamics} (in contrast to statistical mechanics) then arises from the notion of equilibrium states, since these are spontaneously reached but not left without external intervention.

The distinction between the second and the minus first law, and the question where exactly irreversibility is to be located, is very important for providing a microscopic foundation of irreversible processes. \citet[p. 316]{Uffink2001} distinguishes between different meanings of \ZT{irreversibility}, namely \ZT{time-reversal-noninvariance} (there is a process allowed by a theory whose time-reversal is not allowed) and \ZT{irrecoverability} (a transition from a state $A$ to a state $B$ cannot be fully undone). This distinction has also been employed by \citet{Luczak2017}. He claims that minus first and the second law are logically independent statements. The minus first law predicts that systems initially in a nonequilibrium state will irreversibly approach an equilibrium state, where \ZT{irreversible} means \ZT{time-reversal-noninvariant}. The second law, on the other hand, is concerned with transitions between equilibrium states, which can be irreversible in the sense of \ZT{making the initial state irrecoverable}. Statistical mechanics, he concludes, should aim at giving a microscopic underpinning to both the minus first and the second law, which requires two different solutions given that we have two different problems.

Clarifying the location of irreversibility within thermodynamics is thus important in order to decide what precisely we seek an explanation for. Since in this work we are interested in what DDFT can offer, we should therefore ask whether DDFT is related to the second or to the minus first law. As shown by \citet[p. 2348]{Munakata1994}, the deterministic DDFT \eqref{trddft} possesses the H-theorem
\begin{equation}
\frac{\dif F}{\dif t} = \INT{}{}{^3r}\frac{\delta F}{\delta \rho\rt}\frac{\partial}{\partial t}\rho \rt = -\INT{}{}{^3r}\Gamma \rho\rt\bigg(\Nabla\frac{\delta F}{\delta \rho\rt}\bigg)^2 \leq 0.
\end{equation}
In the last step, we have used \cref{trddft} and integration by parts. A minimum is reached for $\delta F / \delta \rho = \mu$ (with chemical potential $\mu$), which is the equilibrium state. Thus, DDFT predicts that nonequilibrium systems spontaneously and monotonously approach an equilibrium state. Moreover, the (local and global) minimum is unique, although local minima can arise in practical applications as a consequence of approximations \cite[p. 8036]{MarconiT1999}. Consequently, DDFT provides a microscopic foundation for the minus first law (for fluids).

\section{\label{two}Second Problem: The Definition of \ZT{Equilibrium}\ and \ZT{Entropy}}
If we wish to explain something, it is, in general, important to have a precise idea about what exactly the thing we need to explain is. In the case of thermodynamics, explaining the increase of entropy or the approach to equilibrium therefore requires that we have an idea about what precisely entropy and equilibrium are. As it turns out, this is not at all clear, which is problematic because different opinions on what equilibrium is lead to different opinions on why and whether equilibrium is approached. Here, I will consider two aspects of the debate concerning thermodynamic equilibrium, which can both be understood better when analyzing the example of DDFT. These are the question whether \ZT{entropy} (and, consequently, \ZT{equilibrium}) is an approximate notion, i.e., whether it should be understood in a fine-grained or a coarse-grained way, and the question whether it is a time-asymmetric notion, i.e., whether fluctuations away from equilibrium are possible.

For the first of these aspects, we consider the argument formulated by \citet{Blatt1959} and \citet{RidderbosR1998} against the coarse-graining procedure, which is based on the famous spin echo experiment performed by \citet{Hahn1950}\footnote{I explain it here in a simplified form, following \citet[pp. 1242-1243]{RidderbosR1998}.}. Spins in a magnetic field are aligned by a strong pulse. The magnetic field causes precession of the spins, and because of small inhomogeneities in the field, the spins precess with different frequencies and are oriented \ZT{randomly} after some time. Hence, they are in a state of coarse-grained equilibrium. Now, a second pulse is applied, which reverses the direction of precession. In consequence, the spins return, after some time, to their initial aligned state. Since the spins were isolated after the second pulse, we appear to have a case of an isolated system spontaneously evolving away from equilibrium, which could be interpreted as a violation of the second law. The explanation for why this happens and why this is \textit{not} a violation of the second law,  \citet[p. 1251]{RidderbosR1998} argue, is that the system was not actually in fine-grained equilibrium. It was only in coarse-grained equilibrium. But since the typical argument for coarse-graining is that we cannot distinguish between fine-grained and coarse-grained equilibrium, and since we here have found a way in which we can, coarse-graining is not a reasonable way of explaining the approach to equilibrium. In particular, if we think of \ZT{equilibrium} as \ZT{coarse-grained equilibrium}, the second part of the spin-echo experiment would constitute a violation of the second law of thermodynamics (which, of course, it is not).

I will get back to the justification of coarse-graining in \cref{three}. What is interesting here is that Ridderbos and Redhead defend a certain characterization of equilibrium as \ZT{fine-grained equilibrium}. They contrast their position with that of \citet[p. 253]{Sklar1995}, who claims that the first part of the spin-echo experiment constitutes \ZT{normal} thermodynamic behaviour. \citet[pp. 1254-1257]{RidderbosR1998}, on the other hand, argue that thermodynamic behaviour would be an approach to \ZT{true} equilibrium, including the loss of correlations. On this perspective, an isolated gas that spreads out in a container would not exhibit an approach to equilibrium, since its distribution would not turn into the uniform equilibrium distribution. This, they conclude, requires external interventions.

But we do not necessarily need to use this definition of equilibrium. Maybe, we are perfectly happy with the weaker characterization of \ZT{equilibrium} as a state in which the values of certain macroscopic quantities are approximately constant \cite[p. 547]{Callender2001}. For example, we macroscopically observe that hot coffee cools down. When we, starting from a microscopic theory, derive a prediction for how the average temperature of the coffee evolves in time, we get, after some approximations, just that: The temperature of the coffee will approach room temperature. These two definitions thus lead to very different meanings of Q4 (\ZT{Why do systems approach equilibrium?}). If we use fine-grained equilibrium as a definition, combined with the very strict reading proposed by Ridderbos and Redhead, then the explanation for why gas expands in an isolated container, or why an isolated cup of coffee cools down, would not be a part of an answer to Q4. Using a definition based on constant macroscopic quantities, on the other hand, we would seek for an explanation of precisely this behaviour.

DDFT is an interesting example for the problem of the approximate nature of thermodynamic equilibrium: As discussed in \cref{ddft}, static DFT allows to find equilibrium states of a many-particle system by minimizing its free energy functional. It is based on a theorem by \citet{Mermin1965} stating that \textit{in equilibrium}, the one-body density uniquely determines the phase-space distribution and thus the free energy. Thereby, DFT provides a significant computational advantage, since the one-body is only a function of the position and not of the phase-space coordinates of all particles. This theorem does, however, no longer hold out of equilibrium. An extension was proven by \citet{RungeG1984} for quantum and by \citet{ChanF2005} for classical time-dependent systems, which shows that in the nonequilibrium case, there is also a functional dependence on the initial state of the system. Consequently, the one-body density does not fully determine the fine-grained free energy for a system that is initially out of equilibrium. Therefore, DDFT describes - strictly speaking - only an approach to coarse-grained equilibrium. If we are in a state that has the same one-body distribution as the equilibrium state, but a different correlation function - i.e., in a state that corresponds to coarse-grained, but not fine-grained equilibrium - then DDFT will not predict that it relaxes, since the right-hand side of \cref{trddft} is zero in this case. Problems of this form are relevant, e.g., for the glass transition \cite[p. 1124]{HeinrichsDMF2004}.

Given the approximate nature of DDFT, a significant amount of research has been devoted to analyzing its shortcomings and potential improvements in the past years. Forces that are not incorporated in DDFT are known as \textit{superadiabatic forces}\footnote{This name is motivated by the fact that DDFT is based on the adiabatic approximation, see \cref{ddft}.}. As shown by \citet{FortinidlHBS2014}, superadiabatic forces can have important effects in systems of Brownian particles. A systematic extension of DDFT that allows to describe them is \textit{power functional theory} (PFT), which was developed by \citet{SchmidtB2013}. This exact formalism is based on a variational principle for the power dissipation, where contributions that go beyond DDFT are incorporated in an \ZT{excess power functional}. An important superadiabatic effect is memory: The Mori-Zwanzig formalism shows that, when the full microscopic description of a system is reduced to the subdynamics of the relevant degrees of freedom, one obtains a dependence on the history of the system \citep[p. 1518]{TreffenstadtS2020}. This dependence is usually neglected in derivations of irreversible dynamical theories, this holds both for the Mori-Zwanzig formalism and for DDFT.

The argument from interventionists outlined above is based on the objection that the history dependence cannot be neglected, since it is responsible for the motion reversal observed in the second part of the spin-echo experiment. Similar effects have also been found in PFT: \citet{TreffenstadtS2020} have recently studied a system of Brownian particles under shear using PFT and Brownian dynamics simulations. Starting from equilibrium, an inhomogeneous shear force field is switched on, such that the system settles to a nonequilibrium steady state. When the shear is switched off again, the system exhibits a global current reversal before approaching equilibrium. This is a consequence of the memory-induced superadiabatic forces, which balance the external forces in steady state and become driving forces after the external forces are switched off. This effect is reminiscent of the spin-echo effect\footnote{I do not claim here that this is the same effect - there are important differences, such as the fact that the spins are a closed quantum system, while the Brownian particles are a dissipative classical system. Nevertheless, the general effect (a system exhibits a motion reversal rather than the monotonous approach to equilibrium simpler models would predict) \textit{is} similar, and given that philosophers of physics tend to be more interested in closed quantum systems, dissipative classical systems offer a lot of unexplored potential for them.}, which indicates that the large amount of work on superadiabatic forces that has emerged in the past years is also relevant for foundational research.

The comparison of fine-grained and coarse-grained equilibrium and entropy leads us to the second aspect if we consider the connection to thermodynamics (as discussed in \cref{one}): If we define \ZT{equilibrium} as \ZT{the state with maximal fine-grained entropy}, then the minus first law is, due to Liouville's theorem stating that the fine-grained entropy is constant, simply wrong (recall that the minus first law is a statement about closed systems). For approaches based on coarse-graining, on the other hand, the minus first law does not pose a problem. \citet[p. 530]{BrownU2001} therefore distinguish in their discussion two different notions of equilibrium: In thermodynamics, the concept of equilibrium is introduced through the minus first law. Here, it is a state that the system spontaneously enters and then remains in. Therefore, equilibrium is by its nature a time-asymmetric notion. In (Boltzmannian) statistical mechanics, on the other hand, equilibrium is the macrostate with the largest entropy or phase-space volume, i.e., the state that can be realized with the largest number of microscopic configurations. This definition is time-symmetric, since deviations are, in principle, possible in both directions of time.

Deterministic DDFT does, as shown in \cref{one}, predict a monotonous approach to an equilibrium state, and thus provides a microscopic basis for the minus first law (for soft matter systems). Once the system has reached a minimum of the free energy, it will stay there forever. This is not the case in the stochastic DDFT \eqref{trddfts}: Here, due to the presence of the noise term, there will still be fluctuations once an equilibrium state is reached. This is more reminiscent of the time-symmetric idea from statistical mechanics, where deviations from equilibrium are allowed. However, both stochastic and deterministic DDFT are microscopic, statistical-mechanical theories. The question whether noise terms should be included in DDFT has led to an intense debate in statistical physics. An important argument against the presence of noise terms, presented by \citet[p. 8034]{MarconiT1999}, is that the free energy functional of DFT already includes all fluctuations such that one would overcount them if they are added to the evolution equation.

An explanation was provided by \citet{ArcherR2004}, who showed that the difference between \cref{trddft} and \cref{trddfts} arises because the density $\rho$ has a different meaning in both theories. As an example, let us compare the derivation of stochastic DDFT by \citet{Dean1996} and the derivation of deterministic DDFT by \citet{MarconiT1999}. \citet{Dean1996} considered the microscopic density
\begin{equation}
\hat{\rho}\rt = \sum_{i=1}^{N}\delta(\vec{r}-\vec{r}_i(t)),
\label{densityoperator}
\end{equation}
where $\delta(\vec{r})$ is the Dirac delta distribution and $N$ is the number of particles. (The density $\hat{\rho}$ is often denoted \ZT{density operator}, and is written with a hat to distinguish it from averaged densities.) He then showed that, if the individual particles obey the Langevin equations \eqref{langevin}, the density \eqref{densityoperator} satisfies the exact evolution equation
\begin{equation}
\frac{\partial}{\partial t}\hat{\rho}\rt=\Gamma \Nabla\cdot\bigg(\hat{\rho}\rt\vec{\nabla}\frac{\delta F[\hat{\rho}]}{\delta\hat{\rho}\rt}\bigg) +\Nabla\cdot\bigg(\sqrt{2\Gamma k_B T\hat{\rho}\rt}\vec{\eta}\rt\bigg),
\label{dk}
\end{equation}
with the free energy
\begin{equation}
F[\hat{\rho}] = \frac{1}{2}\INT{}{}{^3r}\INT{}{}{^3r'}\hat{\rho}\rt U_2(\vec{r}-\vec{r}')\hat{\rho}(\vec{r}',t) + k_BT\INT{}{}{^3r}\hat{\rho}\rt(\ln(\Lambda^3\hat{\rho}\rt)-1), 
\label{deanfreeenergy}
\end{equation}
Here, $U_2(\vec{r})$ is a two-body interaction potential and $\Lambda$ is the (irrelevant) thermal de Broglie wavelength. Notably, the free energy \eqref{deanfreeenergy} is \textit{not} the free energy of DFT, i.e., of equilibrium statistical mechanics. In particular, the interaction term is very complicated and not known in general in DFT, whereas it can be constructed straightforwardly from the interaction potential in \cref{deanfreeenergy}. \citet{MarconiT1999}, on the other hand, studied the ensemble-averaged density
\begin{equation}
\rho\rt = \bigg\langle\sum_{i=1}^{N}\delta(\vec{r}-\vec{r}_i(t))\bigg\rangle,
\end{equation}
where $\braket{\cdot}$ is an average over the microscopic noise. This density can be shown to follow (after an adiabatic approximation, see \cref{ddft}) the deterministic equation \eqref{trddft}. (Another form of stochastic DDFT was obtained by \citet{Kawasaki1994}, this will be discussed in \cref{three}.)

This distinction is of central importance for theoretical soft matter physics, but it is also interesting for the discussion in philosophy of physics: Here, it is (as shown above) a matter of debate to which extent the monotonous approach to equilibrium predicted by thermodynamics is a consequence of approximations, and whether systems will be found to fluctuate out of equilibrium. Comparing the two forms of DDFT discussed here shows that this depends, to a very large extent, on the way we define our observables (here $\rho$) and our thermodynamic functionals (here $F$), in particular on whether they are understood as ensemble averages. This point deserves much more attention in foundational discussions than it typically gets.

It also relates in an interesting way to the distinction between the Gibbs and the Boltzmann approach explained in \cref{theproblem}: As discussed by \citet{Uffink2006}, these framework differ in their conceptions of equilibrium: For Boltzmann, it is a property of a single system. A system that is in equilibrium is not guaranteed to stay there forever. For Gibbs, on the other hand, the idea of equilibrium applies to ensembles, and an ensemble that is in equilibrium will remain there forever (since equilibrium corresponds to a stationary distribution). In DDFT, one has an equilibrium state which the density remains in forever if and only one considers the ensemble-averaged density - this is the case of deterministic DDFT, which is then analogous to the Gibbsian approach. In stochastic DDFT, on the other hand, the system may always fluctuate away from the state that minimizes the free energy, and the density is not to be understood as an ensemble average.

\section{\label{three}Third Problem: The Justification of Coarse-Graining}
The third problem is the question whether and how we can justify coarse-graining. It is an important point here that \ZT{Can we use coarse-graining to explain the approach to equilibrium?} and \ZT{\textit{How} can we use coarse-graining to explain the approach to equilibrium?} are two very different questions. While the second one requires, of course, that the first one has been answered with \ZT{yes}, it has a different content.

A first point we need to make here is that, to not get things mixed up, it is important to distinguish \ZT{coarse-graining}, which is a mathematical technique, from the position of \ZT{the coarse-grainers}, which is the aim of criticism from interventionists. If, in the philosophical literature, someone criticises \ZT{the coarse-grainers}, the criticism is usually aimed at people who use a coarse-grained definition of equilibrium, typically combined with an epistemic justification based on finite measurement resolutions \citep{RidderbosR1998}. This criticism, however, is \textit{not} directed at the procedure of coarse-graining as such (at least it should not be), it is directed at a particular justification of coarse-graining or at a particular definition of \ZT{equilibrium}. The method of coarse-graining, in particular in the form of projection operators, can also be used to formally describe the influence of external interventions, and thus form a basis for interventionism \cite[p. 556]{Robertson2018}. Since \ZT{the position of the coarse-grainers} is just one possible justification of the mathematical procedure of coarse-graining, and since this mathematical procedure is used very successfully throughout physics, one should be very careful with rejecting it altogether, and if one does, this requires a very good justification.

To see why Q3 is a problem on its own, it is helpful to compare the way coarse-graining is used and discussed in physics and in philosophy. Philosophers of physics are interested in coarse-graining because of the role it plays in the explanation of the approach to equilibrium\footnote{I do not claim here that philosophers always believe that coarse-graining is only applicable to thermodynamic irreversibility, the relation of different levels of description is also studied in a more general context \citep{List2019}. Nevertheless, in the philosophy of statistical mechanics, coarse-graining is typically discussed as a possible origin of thermodynamic irreversibility, which differs from the way it is used, e.g., in active matter physics.}. Physicists, on the other hand, also use it in situations where the approach to equilibrium is of no interest or not even expected. DDFT is an excellent case study for this point, which receives little attention in the foundations of physics. It is a coarse-grained theory that, its standard form, describes the irreversible approach to thermodynamic equilibrium. However, modified forms of DDFT that are derived by similar coarse-graining procedures do \textit{not} describe the approach to thermodynamic equilibrium. Hence, interesting insights can be gained by comparing these different forms of DDFT.

A field of application of DDFT where equilibration is not observed is \textit{active matter}. Active particles are particles which use energy in order to create directed motion. A typical example would be a swimming bacterium. Systems consisting of active particles are permanently out of equilibrium, i.e., they do (as long as there is enough energy available) never reach or approach a state of thermodynamic equilibrium. In the description of active particles in statistical mechanics, coarse-graining methods are frequently used to derive a macroscopic field theory based on a microscopic theory of the individual active particles (see, e.g., \citet{BickmannW2020b,BickmannW2020} for such a derivation). The theories derived in this way provide good descriptions for the dynamics of active matter systems.

A DDFT for active particles can be constructed in various ways. \citet{WensinkL2008} described uniaxial\footnote{A particle is uniaxial if it has an axis of continuous rotational symmetry \citep{teVrugtW2020}.} active particles by adding to a DDFT for passive (non-active) nonspherical particles a term accounting for self-propulsion. This was generalized to particles with arbitrary shape by \citet{WittkowskiL2011}. An alternative approach, considered, e.g., by \citet{EnculescuS2011} and \citet{WittmannB2016}, is to construct a DDFT equation of the form
\begin{equation}
\frac{\partial}{\partial t}\rho\rt=\Gamma\Nabla\cdot\bigg(\rho\rt\Nabla\frac{\delta F_{\mathrm{eff}}[\rho]}{\delta\rho\rt}\bigg).
\label{effective}
\end{equation}
Although \cref{effective} looks exactly like \cref{trddft}, it uses an \ZT{effective} free energy $F_{\mathrm{eff}}$ rather than the equilibrium free energy $F$. In the construction of $F_{\mathrm{eff}}$, the external potential acting on the system is modified by the steady-state active force. Since active systems are not in equilibrium, the DDFT \eqref{effective} shows that a coarse-grained theory that describes the approach to a stationary state does not necessarily also describe the approach to thermodynamic equilibrium.

The justification of coarse-graining can then proceed in various ways. One option is to justify the replacement of the fine-grained by a coarse-grained density by our ignorance about microscopic details. \citet[pp. 565-570]{Robertson2018} justifies coarse-graining by its ability to reveal autonomous dynamics on a higher level of description. Notably, both justifications do not require a connection between coarse-graining and equilibrium and are thus applicable more generally. We can also be indifferent about microscopic details of an active system, and active systems also show autonomous dynamics on higher levels of description.

\citet[p. 561]{Robertson2018} has suggested to split the project of justifying coarse-graining into two further sub-problems: We need to justify both why we coarse-grain \textit{at all} and why we coarse-grain \textit{in a particular way}. The answers to these questions might be linked, but do not have to be identical. Up to now, we have been concerned with the first sub-problem. We now turn to problem of why one should coarse-grain \textit{in a particular way}. Again, it is useful to study DDFT.

As discussed in \cref{ddft}, DDFT exists in stochastic and deterministic forms. In \cref{two}, we have compared the deterministic DDFT by \citet{MarconiT1999} to the stochastic DDFT by \citet{Dean1996}, and noted that the former describes the ensemble-averaged density, while the latter is concerned with the microscopic density operator. Another form of stochastic DDFT was derived by \citet{Kawasaki1994}. In this theory, the variable $\rho\rt$ denotes a density after spatial coarse-graining, which is done by dividing space into many small cells. As discussed by \citet{ArcherR2004}, the different definition of the density in Kawasaki's theory is the reason why it is governed by a stochastic rather than a deterministic theory (since the average is taken over space and not over realizations of the noise, the noise does not vanish after averaging). Since the presence or absence of noise terms does, as emphasized in \cref{two}, make a difference for the way equilibrium is approached and for the resulting equilibrium state, the precise way in which coarse-graining is done should be paid attention to when discussing irreversibility.

More generally, the relation between stochastic and deterministic approaches can also be understood from the Mori-Zwanzig formalism, which, as mentioned in \cref{ddft}, allows to derive DDFT. The Mori-Zwanzig formalism exists in various forms, which differ in the form of the projection operator \citep[p. 6345]{Kawasaki2000}: In a \textit{microcanonical} projection operator formalism, the values of the macroscopic variables are specified exactly. When using a \textit{canonical} projection operator, on the other hand, one only specifies the average values of the relevant variables. A canonical projection operator allows to derive deterministic DDFT, with the free energy $F$ being given by the equilibrium free energy of DFT. With a microcanonical projection operator, one can derive the stochastic form of DDFT. The resulting free energy is not identical to the DFT free energy, a connection can be established using a fluctuation renormalization \citep{Kawasaki2006b,teVrugtLW2020}. Philosophical discussions of the Mori-Zwanzig formalism do usually not distinguish between different types of projection operators. However, the example of DDFT shows that this is more than a technical difference, since different projection operators lead to different forms of the dynamic equations that, as shown in \cref{two}, describe the approach to equilibrium in a different way.

\section{\label{four}Fourth Problem: The (Irreversible) Approach to Equilibrium}
While the first three problems have been setting the stage, we now dive more deeply into what is usually thought of as \ZT{the problem of irreversibility}: We wish to explain why systems that are initially out of equilibrium tend to move towards an equilibrium state.

In physics, derivations of irreversible equations from the reversible microscopic physics have a long tradition, starting from Boltzmann's H-theorem (see \citet{Boltzmann1872,BrownUM2009}). \ZT{H-theorem} is, in fact, now a standard name for a proof that a certain theory leads to a monotonous behaviour of (for example) the entropy \citep{AneroE2007,EspanolL2009}. Here is a typical idea of how this can work \cite[pp. 321-323]{North2011}: As discussed above, the entropy is, in statistical mechanics, introduced as a measure for the volume in phase space. Equilibrium is the macrostate with the largest phase space volume.\footnote{I assume here, for the sake of the argument, that these are acceptable definitions of \ZT{entropy} and \ZT{equilibrium}.}. In fact, the phase space volume of the equilibrium state is overwhelmingly larger than that of other states. Hence, if we move into some direction in phase space, it is far more likely that we are going towards equilibrium than away from it (the latter result would correspond to extremely special initial conditions). The approach to equilibrium is therefore explained by statistical considerations. This argument goes back to Boltzmann, and notably, it is a \textit{probabilistic} argument \cite[pp. 185-187]{BrownUM2009}. In the Gibbsian framework, it can be shown that the \textit{coarse-grained} distribution function tends towards a uniform distribution on phase space.

Today, a popular framework for the description of irreversible processes is \textit{stochastic dynamics} \cite[pp. 1038-1063]{Uffink2006}. Here, the laws governing the behaviour of the constituents of a system are assumed not to be deterministic, but stochastic. In particular, a specific type of stochastic processes known as \textit{Markov processes} is employed \citep{Sober2020}. A process is Markovian if it has no memory. This means that (the probability of) what happens in the future does not depend on what has happened in the past. For example, the probability for getting \ZT{heads} in a fair coin toss is always 0.5, no matter how often the coin landed on \ZT{heads} in previous coin tosses. The Markov processes employed in stochastic dynamics have the attractive feature that they tend towards an equilibrium distribution after a while, so if we describe a system in statistical mechanics by Markovian dynamical equations, then chances are good that we will obtain a description of the irreversible approach to equilibrium. (Mathematically, the description in terms of Markov processes corresponds to using so-called \ZT{Master equations}, which can be obtained as an approximation of the actual microscopic dynamics within the Mori-Zwanzig formalism \cite[pp. 57-68]{Zeh1989}.)

The question is, of course, why we should be justified to model a system using stochastic Markov processes given that the actual dynamics is governed by Hamilton's equations, which are deterministic. Three main viewpoints can be found in the literature \cite[pp. 1038-1039]{Uffink2006}. The first one is an appeal to coarse-graining \citep{vanKampen2002}: We ignore microscopic details of the system that we cannot measure anyway and focus on macrostates. Then, we can find transition probabilities for the change from one macrostate to another one and thus obtain a stochastic process. Another option is interventionism \citep{Blatt1959,RidderbosR1998}: The system we describe is not actually closed, but subject to external influences. Since these are not known, our description of the system is stochastic. Finally, some authors have advocated being agnostic about the origin of stochasticity and simply treating stochastic dynamics as fundamental \citep{Streater2009}.

Importantly, these approaches are not necessarily in contradiction: Recall that on the interventionist's definition, the facts that gas in an isolated container spreads out, that a hot cup of coffee in an isolated room aquires room temperature, or that the spins go out of phase in the first part of the spin echo experiment do not constitute instances of an approach to equilibrium. Nevertheless, they do not deny that these things happen. Now let us have a look at how these effects are explained in the physics literature, where coarse-graining is employed. An illustration used by \citet[p. 14]{DonevFVE2014} is a system consisting of many oscillators that, after some time, will go out of phase. Every individual oscillator follows a time-reversible law, yet the average amplitude will decay due to dephasing. This, remarkably, is exactly the explanation that \citet[p. 1242]{RidderbosR1998} use for what happens in the first part of the spin-echo experiment - the spins oscillate and go out of phase. Thus, there is no difference between the physical mechanisms by which interventionists explain the spread out of an isolated gas and the mechanisms by which a physicist using coarse-graining would explain the same effect. The major difference is that for an interventionist, \ZT{approach to equilibrium} means something different, such that different mechanisms (namely external interventions) are required to explain it. Therefore, interventionists can still agree that the coarse-grained density of a closed system will tend to a uniform distribution, they will just not classify this as an \ZT{approach to equilibrium}.

Again, DDFT provides an instructive example. As discussed in \cref{ddft}, one can distinguish between DDFT for colloidal fluids and for atomic fluids. In the case of atomic fluids, the microscopic dynamics is given by Hamilton's equations, and is therefore reversible. Hence, a derivation of the irreversible DDFT equation \eqref{trddft} requires a coarse-graining procedure. A good example is the derivation of DDFT from the Mori-Zwanzig formalism by \citet{EspanolL2009}: Starting from the reversible Hamiltonian dynamics, one obtains an exact equation of motion for the relevant variable (here the one-body density) that involves a memory term. It is then assumed that the one-body density is a slow variable (more precisely: that it changes in time much more slowly than the current correlation function), such that memory effects can be neglected. This leads to the irreversible DDFT equation\footnote{Strictly speaking, it leads to a non-local generalization of \cref{trddft}, that has to be made local by the additional assumption that the system is a dilute suspension. This, however, is a technical problem that is not relevant for the present discussion.} \eqref{trddft}. Thus, it is a typical example for the type of derivation \ZT{coarse-grainers} have in mind when discussing irreversibility: We start from a reversible system, coarse-grain it, make the approximation that the relevant part of the dynamics is Markovian, and arrive at an irreversible law.

However, DDFT is also (in fact: more often) applied to colloidal fluids. Here, the microscopic dynamics is given by the Langevin equations \eqref{langevin}. Since these are already asymmetric in time \cite[p. 405]{Luczak2016}, coarse-graining is not required to obtain irreversibility. Stochasticity here comes into play through the noise term in \cref{langevin}. As shown by \citet{Dean1996}, one can derive from \cref{langevin} an exact irreversible DDFT equation of the form \eqref{trddfts}, which describes an approach to equilibrium. (The additional approximations employed by \citet{MarconiT1999} are required to obtain a closed equation for the ensemble-averaged one-body density.) This is a typical example of a derivation that interventionists think of when discussing irreversibility: The colloidal system approaches equilibrium because it is - due to collisions of the colloids with the solvent particles, which act as a heat bath and provide noise - not a closed system. Consequently, both main approaches to stochastic dynamics - coarse-graining and interventionism - are required to understand modern nonequilibrium statistical mechanics, since both play a role in DDFT.

What is, moreover, interesting is that there is not necessarily a disagreement between these two positions: The irreversible Langevin equations \eqref{langevin} can be derived by coarse-graining the full (reversible) microscopic equations of motion for the system consisting of both colloidal and solvent particles. What interventionists and coarse-grainers would disagree on is whether an atomic fluid or a closed system consisting of colloids and solvents can be said to approach equilibrium. However, as shown above, this is merely a consequence of the way \ZT{equilibrium} is defined in these two approaches.

\section{\label{five}Fifth Problem: The Arrow of Time}
We have now reached the final problem, which is concerned with the \textit{arrow of time}. Let us assume we have found some explanation for why, if we go forwards in time, the systems are observed to approach equilibrium, i.e., that we have answered question four. We now need to find out why, despite the symmetry of the fundamental laws of physics, this answer cannot be applied to the past.

The distinction between Q4 and Q5 allows us to understand why, as mentioned in \cref{theproblem}, philosophers and physicists concerned with the origin of irreversibility are discussing it based on considerations about the big bang and cosmology, while researchers working on nonequilibrium statistical mechanics are successful without ever spending a thought on these issues \citep{Penrose1994}. The methods employed in statistical physics - coarse-graining and Markovian approximations - allow for a solution of Q4. This is sufficient if one takes the temporal asymmetry as given, which is why the early universe plays no role in the everyday life of a statistical physicist. Philosophers (and physicists) who are interested in the source of this asymmetry, on the other hand, need to address Q5, which requires a different type of answer.

Recall how the standard solution discussed in \cref{theproblem} works: We start from a reversible theory describing the dynamics of the density  and then perform a coarse-graining, in which we replace the fine-grained by the coarse-grained density, allowing to define a coarse-grained entropy. The dynamics of the coarse-grained density then leads to an increase of the coarse-grained entropy. Unfortunately, the original microdynamics we started with was time-reversal symmetric, i.e., it does not know a difference between past and future. Hence, we could have simply decided to perform the coarse-graining in the other direction of time, which would have been mathematically just as well justified as the original derivation. This \ZT{backwards coarse-graining} would then lead to a theory in which the entropy decreases in the future and increases in the past, in strong contradiction to every observation \cite[p. 15]{Wallace2011}.

\citet[pp. 115-119]{Albert2000} has explained this problem as a problem of inference: Q4 is, as discussed in \cref{four}, frequently solved by arguing that an evolution towards equilibrium is more likely on statistical grounds. Albert approaches this from an epistemological perspective: In principle, he argues, we have two ways of obtaining information. \textit{Prediction} means that we, from the known laws of physics and the present state of the world as an initial condition, infer what will happen in the future. Doing the same for the past is called \textit{retrodiction}. The alternative is to use \textit{records}, from which we obtain most of our knowledge about the past. The question is now, Albert argues, why we should have sources of information other than prediction or retrodiction given that we only have direct reliable empirical information about the present state of the world. Unfortunately, if we use classical mechanics combined with statistical considerations about phase space to retrodict what happened in the past based on our knowledge of the present, we will conclude - by the line of argument employed in \cref{four} - that the past had a higher entropy than the present. Hence, most of our records would most likely be wrong. Even worse, given that we only believe in the laws of mechanics because of experiments that we have records of, we could infer from classical mechanics that classical mechanics is, presumably, also wrong. Thus, statistical mechanics has brought us into a position of scepticism. To avoid this, he introduces an additional postulate, the \textit{past hypothesis}: The entropy of the initial state of the universe was very low. If we then use, as a basis for our inference, not only the laws of mechanics and the statistics of phase space, but in addition also the past hypothesis, then we can believe that our records of the past are, in general true. Thus, we have avoided Albert's problem of scepticism.

\citet{Wallace2011} has suggested an alternative formulation\footnote{See \citet{Brown2017} for a comparison of Albert's and Wallace's formulation.}, the \textit{simple past hypothesis}: He defends the position that coarse-graining is justified if the density is forwards compatible, i.e., if it does not affect the predictions for the macroscopic variables. This is satisfied if the density is \ZT{simple}.\footnote{See \citet{Wallace2011} for a more mathematical discussion.} Unfortunately, a simple distribution is also backwards compatible, i.e., we can perform a coarse-graining in the other direction of time, predicting an (unphysical) increase of entropy in the past. To avoid this, and therefore to answer Q5, we assume that the the initial state of the universe was simple. In this case, we cannot perform a problematic backwards coarse-graining, since the initial state of the universe has no past.

DDFT is an interesting example also for this problem. As mentioned in \cref{four}, it is possible in the Mori-Zwanzig formalism to derive the irreversible DDFT equation from the reversible laws of classical mechanics, a theory where the entropy always increases. However, this only answers Q4 - depending on the choice of initial conditions, it is also possible in the Mori-Zwanzig formalism to derive a theory in which entropy always \textit{decreases} \cite[p. 67]{Zeh1989}. This depends on the choice of initial conditions: To obtain a closed equation of motion in which the entropy increases, we have to assume that the irrelevant part of the phase-space density vanishes at the initial time \citep[p. 292]{Wallace2015}. This assumption is therefore also required for derivations of DDFT in the Mori-Zwanzig formalism. Unfortunately, this is not always made clear in such derivations. An explicit discussion of this hypothesis and its role in DDFT was given by \citet{Yoshimori2005}.

When setting up the initial condition in a different way, the irreversible dynamics would apply backwards in time. This possibility is, in the philosophy of statistical mechanics, usually dismissed as an unphysical artefact that one has to get rid of. However, it is notable that a \ZT{backwards coarse-grained theory} is sometimes actually employed in statistical physics \citep{Lutsko2019}: In the theory of nucleation, where particles in a fluid aggregate to a nucleus that then grows leading to solidification, one is interested in calculating the \ZT{most likely path} along which nucleation occurs. For nucleation, the system has to overcome an energy barrier. The most likely path can be obtained by starting at the top of the energy barrier, evolving the system forwards in time by the (coarse-grained and irreversible) DDFT equation towards the final state and backwards in time (!) towards a the initial state which had a lower free energy. \citet[p. 2]{Lutsko2011}, who introduced this procedure, emphasizes that it is only a mathematical trick. Nevertheless, it is an interesting observation that this is mathematically possible and occasionally useful, and it is an interesting question why this is only a mathematical trick. 

\section{\label{conclusion}Conclusion}
In summary, I have shown that the study of dynamical density functional theory (DDFT) provides interesting and important insights for the philosophy of statistical mechanics. This has been demonstrated for each of the five \ZT{problems of irreversibility} discussed here: DDFT provides a microscopic basis for the minus first law (Q1), it reveals the importance of taking into account memory effects and of distinguishing between various types of free energy (Q2), it illustrates the subtle relation between irreversibility and coarse-graining as well as the fact that different forms of coarse-graining lead to different types of irreversible theories (Q3), it shows that both a coarse-grained description and external interventions are possible sources of irreversible behavior (Q4), and it reveals that it can be useful to apply coarse-grained theories backwards in time (Q5). Consequently, for philosophers of physics who wish their positions to incorporate results of modern research on nonequilibrium systems, DDFT is a useful starting point for connecting foundational work based on the Mori-Zwanzig formalism to applied research on soft condensed matter systems - a connection that can be expected to be beneficial for both fields.
\section*{Acknowledgements}
I am very grateful to my philosophy supervisor Paul N\"ager for many interesting discussions that helped me in developing the ideas presented here, and for continuous support in writing this article. In addition, I would like to thank Ulrich Krohs, Oliver Robert Scholz and the other participants of the thesis colloquium for useful comments on a preliminary version of this work. I further acknowledge interesting discussions with various members of the M\"unsteranian \ZT{Zentrum f\"ur Wissenschaftstheorie} (Centre for Philosophy of Science). In particular, I am grateful for helpful feedback from Markus Seidel, Gernot M\"unster, J\"org Friedrich, Renzo Kapust and the other participants of the \ZT{Arbeitskreis Wissenschaftstheorie} (Philosophy of Science discussion group). Moreover, I thank Berthold te Vr\"ugt for a critical reading of the manuscript. Finally, I thank my physics supervisor Raphael Wittkowski for supporting this project and for introducing me to the relevant physics.

\section*{Funding}
This work is funded by the Deutsche Forschungsgemeinschaft (DFG, German Research Foundation) -- grant number WI 4170/3-1. I also wish to thank the Studienstiftung des deutschen Volkes for financial support.
\bibliography{refsphil}{}

\begin{thebibliography}{92}
\expandafter\ifx\csname natexlab\endcsname\relax\def\natexlab#1{#1}\fi
\providecommand{\url}[1]{\texttt{#1}}
\providecommand{\href}[2]{#2}
\providecommand{\path}[1]{#1}
\providecommand{\DOIprefix}{doi:}
\providecommand{\ArXivprefix}{arXiv:}
\providecommand{\URLprefix}{URL: }
\providecommand{\Pubmedprefix}{pmid:}
\providecommand{\doi}[1]{\href{http://dx.doi.org/#1}{\path{#1}}}
\providecommand{\Pubmed}[1]{\href{pmid:#1}{\path{#1}}}
\providecommand{\bibinfo}[2]{#2}
\ifx\xfnm\relax \def\xfnm[#1]{\unskip,\space#1}\fi
\bibitem[{Al-Saedi et~al.(2018)Al-Saedi, Archer and Ward}]{AlSaediHAW2018}
\bibinfo{author}{Al-Saedi, H.M.}, \bibinfo{author}{Archer, A.J.},
  \bibinfo{author}{Ward, J.}, \bibinfo{year}{2018}.
\newblock \bibinfo{title}{Dynamical density-functional-theory-based modeling of
  tissue dynamics: application to tumor growth}.
\newblock \bibinfo{journal}{Physical Review E} \bibinfo{volume}{98},
  \bibinfo{pages}{022407}.
\bibitem[{Albert(2000)}]{Albert2000}
\bibinfo{author}{Albert, D.Z.}, \bibinfo{year}{2000}.
\newblock \bibinfo{title}{Time and Chance}.
\newblock \bibinfo{publisher}{Harvard University Press},
  \bibinfo{address}{Cambridge (Massachusetts)}.
\bibitem[{Anero and Espa{\~n}ol(2007)}]{AneroE2007}
\bibinfo{author}{Anero, J.G.}, \bibinfo{author}{Espa{\~n}ol, P.},
  \bibinfo{year}{2007}.
\newblock \bibinfo{title}{Dynamic {B}oltzmann free-energy functional theory}.
\newblock \bibinfo{journal}{Europhysics Letters} \bibinfo{volume}{78},
  \bibinfo{pages}{50005}.
\bibitem[{Angioletti-Uberti et~al.(2018)Angioletti-Uberti, Ballauff and
  Dzubiella}]{AngiolettiBD2018}
\bibinfo{author}{Angioletti-Uberti, S.}, \bibinfo{author}{Ballauff, M.},
  \bibinfo{author}{Dzubiella, J.}, \bibinfo{year}{2018}.
\newblock \bibinfo{title}{Competitive adsorption of multiple proteins to
  nanoparticles: the {V}roman effect revisited}.
\newblock \bibinfo{journal}{Molecular Physics} \bibinfo{volume}{116},
  \bibinfo{pages}{3154--3163}.
\bibitem[{Archer and Evans(2004)}]{ArcherE2004}
\bibinfo{author}{Archer, A.J.}, \bibinfo{author}{Evans, R.},
  \bibinfo{year}{2004}.
\newblock \bibinfo{title}{Dynamical density functional theory and its
  application to spinodal decomposition}.
\newblock \bibinfo{journal}{Journal of Chemical Physics} \bibinfo{volume}{121},
  \bibinfo{pages}{4246--4254}.
\bibitem[{Archer and Rauscher(2004)}]{ArcherR2004}
\bibinfo{author}{Archer, A.J.}, \bibinfo{author}{Rauscher, M.},
  \bibinfo{year}{2004}.
\newblock \bibinfo{title}{Dynamical density functional theory for interacting
  {B}rownian particles: stochastic or deterministic?}
\newblock \bibinfo{journal}{Journal of Physics A: Mathematical and General}
  \bibinfo{volume}{37}, \bibinfo{pages}{9325}.
\bibitem[{Babel et~al.(2018)Babel, Eikerling and L\"owen}]{BabelEL2018}
\bibinfo{author}{Babel, S.}, \bibinfo{author}{Eikerling, M.},
  \bibinfo{author}{L\"owen, H.}, \bibinfo{year}{2018}.
\newblock \bibinfo{title}{Impedance resonance in narrow confinement}.
\newblock \bibinfo{journal}{Journal of Physical Chemistry C}
  \bibinfo{volume}{122}, \bibinfo{pages}{21724--21734}.
\bibitem[{Bickmann and Wittkowski(2020a)}]{BickmannW2020b}
\bibinfo{author}{Bickmann, J.}, \bibinfo{author}{Wittkowski, R.},
  \bibinfo{year}{2020}a.
\newblock \bibinfo{title}{Collective dynamics of active {B}rownian particles in
  three spatial dimensions: a predictive field theory}.
\newblock \bibinfo{journal}{Physical Review Research} \bibinfo{volume}{2},
  \bibinfo{pages}{033241}.
\bibitem[{Bickmann and Wittkowski(2020b)}]{BickmannW2020}
\bibinfo{author}{Bickmann, J.}, \bibinfo{author}{Wittkowski, R.},
  \bibinfo{year}{2020}b.
\newblock \bibinfo{title}{Predictive local field theory for interacting active
  {B}rownian spheres in two spatial dimensions}.
\newblock \bibinfo{journal}{Journal of Physics: Condensed Matter}
  \bibinfo{volume}{32}, \bibinfo{pages}{214001}.
\bibitem[{Blatt(1959)}]{Blatt1959}
\bibinfo{author}{Blatt, J.M.}, \bibinfo{year}{1959}.
\newblock \bibinfo{title}{An alternative approach to the ergodic problem}.
\newblock \bibinfo{journal}{Progress of Theoretical Physics}
  \bibinfo{volume}{22}, \bibinfo{pages}{745--756}.
\bibitem[{Boltzmann(1872)}]{Boltzmann1872}
\bibinfo{author}{Boltzmann, L.}, \bibinfo{year}{1872}.
\newblock \bibinfo{title}{Weitere {S}tudien \"uber das {W}\"armegleichgewicht
  unter {G}asmolek\"ulen}.
\newblock \bibinfo{journal}{Sitzungberichte der Akademie der Wissenschaften zu
  Wien, mathematischnaturwissenschaftliche Klasse, 66} ,
  \bibinfo{pages}{275--370}.
\bibitem[{Brown(2017)}]{Brown2017}
\bibinfo{author}{Brown, H.R.}, \bibinfo{year}{2017}.
\newblock \bibinfo{title}{Once and for all: the curious role of probability in
  the past hypothesis}.
\newblock \bibinfo{journal}{preprint, available at
  http://philsci-archive.pitt.edu/13008/} .
\bibitem[{Brown et~al.(2009)Brown, Myrvold and Uffink}]{BrownUM2009}
\bibinfo{author}{Brown, H.R.}, \bibinfo{author}{Myrvold, W.},
  \bibinfo{author}{Uffink, J.}, \bibinfo{year}{2009}.
\newblock \bibinfo{title}{Boltzmann's {H-}theorem, its discontents, and the
  birth of statistical mechanics}.
\newblock \bibinfo{journal}{Studies in History and Philosophy of Modern
  Physics} \bibinfo{volume}{40}, \bibinfo{pages}{174--191}.
\bibitem[{Brown and Uffink(2001)}]{BrownU2001}
\bibinfo{author}{Brown, H.R.}, \bibinfo{author}{Uffink, J.},
  \bibinfo{year}{2001}.
\newblock \bibinfo{title}{The origins of time-asymmetry in thermodynamics: The
  minus first law}.
\newblock \bibinfo{journal}{Studies in History and Philosophy of Modern
  Physics} \bibinfo{volume}{32}, \bibinfo{pages}{525 -- 538}.
\bibitem[{Callender(1999)}]{Callender1999}
\bibinfo{author}{Callender, C.}, \bibinfo{year}{1999}.
\newblock \bibinfo{title}{Reducing thermodynamics to statistical mechanics: The
  case of entropy}.
\newblock \bibinfo{journal}{Journal of Philosophy} \bibinfo{volume}{96},
  \bibinfo{pages}{348--373}.
\bibitem[{Callender(2001)}]{Callender2001}
\bibinfo{author}{Callender, C.}, \bibinfo{year}{2001}.
\newblock \bibinfo{title}{Taking thermodynamics too seriously}.
\newblock \bibinfo{journal}{Studies in History and Philosophy of Modern
  Physics} \bibinfo{volume}{32}, \bibinfo{pages}{539--553}.
\bibitem[{Callender(2021)}]{Callender2016}
\bibinfo{author}{Callender, C.}, \bibinfo{year}{2021}.
\newblock \bibinfo{title}{Thermodynamic Asymmetry in Time}.
  \bibinfo{edition}{spring 2021} ed.. \bibinfo{publisher}{The Stanford
  Encyclopedia of Philosophy, Metaphysics Research Lab, Stanford University}.
\bibitem[{Chan and Finken(2005)}]{ChanF2005}
\bibinfo{author}{Chan, G.K.L.}, \bibinfo{author}{Finken, R.},
  \bibinfo{year}{2005}.
\newblock \bibinfo{title}{Time-dependent density functional theory of classical
  fluids}.
\newblock \bibinfo{journal}{Physical Review Letters} \bibinfo{volume}{94},
  \bibinfo{pages}{183001}.
\bibitem[{Dean(1996)}]{Dean1996}
\bibinfo{author}{Dean, D.S.}, \bibinfo{year}{1996}.
\newblock \bibinfo{title}{Langevin equation for the density of a system of
  interacting {Langevin} processes}.
\newblock \bibinfo{journal}{Journal of Physics A: Mathematical and General}
  \bibinfo{volume}{29}, \bibinfo{pages}{L613--L617}.
\bibitem[{Diaw and Murillo(2015)}]{DiawM2015}
\bibinfo{author}{Diaw, A.}, \bibinfo{author}{Murillo, M.S.},
  \bibinfo{year}{2015}.
\newblock \bibinfo{title}{Generalized hydrodynamics model for strongly coupled
  plasmas}.
\newblock \bibinfo{journal}{Physical Review E} \bibinfo{volume}{92},
  \bibinfo{pages}{013107}.
\bibitem[{Diaw and Murillo(2016)}]{DiawM2016}
\bibinfo{author}{Diaw, A.}, \bibinfo{author}{Murillo, M.S.},
  \bibinfo{year}{2016}.
\newblock \bibinfo{title}{A dynamic density functional theory approach to
  diffusion in white dwarfs and neutron star envelopes}.
\newblock \bibinfo{journal}{Astrophysical Journal} \bibinfo{volume}{829},
  \bibinfo{pages}{16}.
\bibitem[{Donev et~al.(2014)Donev, Fai and Vanden-Eijnden}]{DonevFVE2014}
\bibinfo{author}{Donev, A.}, \bibinfo{author}{Fai, T.G.},
  \bibinfo{author}{Vanden-Eijnden, E.}, \bibinfo{year}{2014}.
\newblock \bibinfo{title}{A reversible mesoscopic model of diffusion in
  liquids: from giant fluctuations to {F}ick’s law}.
\newblock \bibinfo{journal}{Journal of Statistical Mechanics: Theory and
  Experiment} \bibinfo{volume}{2014}, \bibinfo{pages}{P04004}.
\bibitem[{{Enculescu} and {Stark}(2011)}]{EnculescuS2011}
\bibinfo{author}{{Enculescu}, M.}, \bibinfo{author}{{Stark}, H.},
  \bibinfo{year}{2011}.
\newblock \bibinfo{title}{Active colloidal suspensions exhibit polar order
  under gravity}.
\newblock \bibinfo{journal}{Physical Review Letters} \bibinfo{volume}{107},
  \bibinfo{pages}{058301}.
\bibitem[{Espa\~{n}ol and V{\'a}zquez(2002)}]{EspanolV2002}
\bibinfo{author}{Espa\~{n}ol, P.}, \bibinfo{author}{V{\'a}zquez, F.},
  \bibinfo{year}{2002}.
\newblock \bibinfo{title}{Coarse graining from coarse-grained descriptions}.
\newblock \bibinfo{journal}{Philosophical Transactions of the Royal Society of
  London. Series A: Mathematical, Physical and Engineering Sciences}
  \bibinfo{volume}{360}, \bibinfo{pages}{383--394}.
\bibitem[{Espa{\~n}ol and L\"owen(2009)}]{EspanolL2009}
\bibinfo{author}{Espa{\~n}ol, P.}, \bibinfo{author}{L\"owen, H.},
  \bibinfo{year}{2009}.
\newblock \bibinfo{title}{Derivation of dynamical density functional theory
  using the projection operator technique}.
\newblock \bibinfo{journal}{Journal of Chemical Physics} \bibinfo{volume}{131},
  \bibinfo{pages}{244101}.
\bibitem[{Evans(1979)}]{Evans1979}
\bibinfo{author}{Evans, R.}, \bibinfo{year}{1979}.
\newblock \bibinfo{title}{The nature of the liquid-vapour interface and other
  topics in the statistical mechanics of non-uniform, classical fluids}.
\newblock \bibinfo{journal}{Advances in Physics} \bibinfo{volume}{28},
  \bibinfo{pages}{143--200}.
\bibitem[{Forster(1974)}]{Forster1974}
\bibinfo{author}{Forster, D.}, \bibinfo{year}{1974}.
\newblock \bibinfo{title}{Hydrodynamics and correlation functions in ordered
  systems: nematic liquid crystals}.
\newblock \bibinfo{journal}{Annals of Physics} \bibinfo{volume}{84},
  \bibinfo{pages}{505--534}.
\bibitem[{Fortini et~al.(2014)Fortini, {de las Heras}, Brader and
  Schmidt}]{FortinidlHBS2014}
\bibinfo{author}{Fortini, A.}, \bibinfo{author}{{de las Heras}, D.},
  \bibinfo{author}{Brader, J.M.}, \bibinfo{author}{Schmidt, M.},
  \bibinfo{year}{2014}.
\newblock \bibinfo{title}{Superadiabatic forces in {B}rownian many-body
  dynamics}.
\newblock \bibinfo{journal}{Physical Review Letters} \bibinfo{volume}{113},
  \bibinfo{pages}{167801}.
\bibitem[{Fraaije(1993)}]{Fraaije1993}
\bibinfo{author}{Fraaije, J.G.E.M.}, \bibinfo{year}{1993}.
\newblock \bibinfo{title}{Dynamic density functional theory for microphase
  separation kinetics of block copolymer melts}.
\newblock \bibinfo{journal}{Journal of Chemical Physics} \bibinfo{volume}{99},
  \bibinfo{pages}{9202--9212}.
\bibitem[{Frigg(2008)}]{Frigg2008}
\bibinfo{author}{Frigg, R.}, \bibinfo{year}{2008}.
\newblock \bibinfo{title}{A field guide to recent work on the foundations of
  statistical mechanics}, in: \bibinfo{editor}{Rickles, D.} (Ed.),
  \bibinfo{booktitle}{The Ashgate Companion to Contemporary Philosophy of
  Physics}. \bibinfo{publisher}{Ashgate}, pp. \bibinfo{pages}{99--196}.
\bibitem[{Frigg and Werndl(2021)}]{FriggW2020}
\bibinfo{author}{Frigg, R.}, \bibinfo{author}{Werndl, C.},
  \bibinfo{year}{2021}.
\newblock \bibinfo{title}{Can somebody please say what {G}ibbsian statistical
  mechanics says?}
\newblock \bibinfo{journal}{British Journal for the Philosophy of Science}
  \bibinfo{volume}{72}, \bibinfo{pages}{105--129}.
\bibitem[{Gibbs(1902)}]{Gibbs1902}
\bibinfo{author}{Gibbs, J.W.}, \bibinfo{year}{1902}.
\newblock \bibinfo{title}{Elementary principles in statistical mechanics:
  developed with especial reference to the rational foundation of
  thermodynamics}.
\newblock \bibinfo{publisher}{C. Scribner's sons}.
\newblock \bibinfo{note}{Available at
  http://eremita.di.uminho.pt/gutenberg/5/0/9/9/50992/50992-pdf.pdf}.
\bibitem[{Grabert(1978)}]{Grabert1978}
\bibinfo{author}{Grabert, H.}, \bibinfo{year}{1978}.
\newblock \bibinfo{title}{Nonlinear transport and dynamics of fluctuations}.
\newblock \bibinfo{journal}{Journal of Statistical Physics}
  \bibinfo{volume}{19}, \bibinfo{pages}{479--497}.
\bibitem[{Grabert(1982)}]{Grabert1982}
\bibinfo{author}{Grabert, H.}, \bibinfo{year}{1982}.
\newblock \bibinfo{title}{Projection Operator Techniques in Nonequilibrium
  Statistical Mechanics}. volume~\bibinfo{volume}{95} of
  \textit{\bibinfo{series}{Springer Tracts in Modern Physics}}.
\newblock \bibinfo{edition}{1} ed., \bibinfo{publisher}{Springer-Verlag},
  \bibinfo{address}{Berlin}.
\bibitem[{Hahn(1950)}]{Hahn1950}
\bibinfo{author}{Hahn, E.L.}, \bibinfo{year}{1950}.
\newblock \bibinfo{title}{Spin echoes}.
\newblock \bibinfo{journal}{Physical Review} \bibinfo{volume}{80},
  \bibinfo{pages}{580}.
\bibitem[{Heinrichs et~al.(2004)Heinrichs, Dieterich, Maass and
  Frisch}]{HeinrichsDMF2004}
\bibinfo{author}{Heinrichs, S.}, \bibinfo{author}{Dieterich, W.},
  \bibinfo{author}{Maass, P.}, \bibinfo{author}{Frisch, H.},
  \bibinfo{year}{2004}.
\newblock \bibinfo{title}{Static and time dependent density functional theory
  with internal degrees of freedom: Merits and limitations demonstrated for the
  {P}otts model}.
\newblock \bibinfo{journal}{Journal of Statistical Physics}
  \bibinfo{volume}{114}, \bibinfo{pages}{1115--1125}.
\bibitem[{{Hohenberg} and {Kohn}(1964)}]{HohenbergK1964}
\bibinfo{author}{{Hohenberg}, P.}, \bibinfo{author}{{Kohn}, W.},
  \bibinfo{year}{1964}.
\newblock \bibinfo{title}{Inhomogeneous electron gas}.
\newblock \bibinfo{journal}{Physical Review} \bibinfo{volume}{136},
  \bibinfo{pages}{864--871}.
\bibitem[{van Kampen(2002)}]{vanKampen2002}
\bibinfo{author}{van Kampen, N.G.}, \bibinfo{year}{2002}.
\newblock \bibinfo{title}{The road from molecules to {O}nsager}.
\newblock \bibinfo{journal}{Journal of Statistical Physics}
  \bibinfo{volume}{109}, \bibinfo{pages}{471--481}.
\bibitem[{Kawasaki(1994)}]{Kawasaki1994}
\bibinfo{author}{Kawasaki, K.}, \bibinfo{year}{1994}.
\newblock \bibinfo{title}{Stochastic model of slow dynamics in supercooled
  liquids and dense colloidal suspensions}.
\newblock \bibinfo{journal}{Physica A: Statistical Mechanics and its
  Applications} \bibinfo{volume}{208}, \bibinfo{pages}{35--64}.
\bibitem[{Kawasaki(2000)}]{Kawasaki2000}
\bibinfo{author}{Kawasaki, K.}, \bibinfo{year}{2000}.
\newblock \bibinfo{title}{Theoretical methods dealing with slow dynamics}.
\newblock \bibinfo{journal}{Journal of Physics: Condensed Matter}
  \bibinfo{volume}{12}, \bibinfo{pages}{6343--6351}.
\bibitem[{Kawasaki(2006)}]{Kawasaki2006b}
\bibinfo{author}{Kawasaki, K.}, \bibinfo{year}{2006}.
\newblock \bibinfo{title}{Interpolation of stochastic and deterministic reduced
  dynamics}.
\newblock \bibinfo{journal}{Physica A: Statistical Mechanics and its
  Applications} \bibinfo{volume}{362}, \bibinfo{pages}{249--260}.
\bibitem[{Kawasaki and Gunton(1973)}]{KawasakiG1973}
\bibinfo{author}{Kawasaki, K.}, \bibinfo{author}{Gunton, J.D.},
  \bibinfo{year}{1973}.
\newblock \bibinfo{title}{Theory of nonlinear transport processes: nonlinear
  shear viscosity and normal stress effects}.
\newblock \bibinfo{journal}{Physical Review A} \bibinfo{volume}{8},
  \bibinfo{pages}{2048--2064}.
\bibitem[{List(2019)}]{List2019}
\bibinfo{author}{List, C.}, \bibinfo{year}{2019}.
\newblock \bibinfo{title}{Levels: descriptive, explanatory, and ontological}.
\newblock \bibinfo{journal}{No{\^u}s} \bibinfo{volume}{53},
  \bibinfo{pages}{852--883}.
\bibitem[{Liu and Liu(2020)}]{LiuL2020}
\bibinfo{author}{Liu, Y.}, \bibinfo{author}{Liu, H.}, \bibinfo{year}{2020}.
\newblock \bibinfo{title}{Development of reaction-diffusion {DFT} and its
  application to catalytic oxidation of {NO} in porous materials}.
\newblock \bibinfo{journal}{AIChE Journal} \bibinfo{volume}{66},
  \bibinfo{pages}{e16824}.
\bibitem[{Luczak(2016)}]{Luczak2016}
\bibinfo{author}{Luczak, J.}, \bibinfo{year}{2016}.
\newblock \bibinfo{title}{On how to approach the approach to equilibrium}.
\newblock \bibinfo{journal}{Philosophy of Science} \bibinfo{volume}{83},
  \bibinfo{pages}{393--411}.
\bibitem[{Luczak(2017)}]{Luczak2017}
\bibinfo{author}{Luczak, J.}, \bibinfo{year}{2017}.
\newblock \bibinfo{title}{How many aims are we aiming at?}
\newblock \bibinfo{journal}{Analysis} \bibinfo{volume}{78},
  \bibinfo{pages}{244--254}.
\bibitem[{Lutsko(2011)}]{Lutsko2011}
\bibinfo{author}{Lutsko, J.F.}, \bibinfo{year}{2011}.
\newblock \bibinfo{title}{Communication: A dynamical theory of homogeneous
  nucleation for colloids and macromolecules}.
\newblock \bibinfo{journal}{Journal of Chemical Physics} \bibinfo{volume}{135},
  \bibinfo{pages}{161101}.
\bibitem[{Lutsko(2019)}]{Lutsko2019}
\bibinfo{author}{Lutsko, J.F.}, \bibinfo{year}{2019}.
\newblock \bibinfo{title}{How crystals form: A theory of nucleation pathways}.
\newblock \bibinfo{journal}{Science Advances} \bibinfo{volume}{5}.
\newblock \bibinfo{note}{Eaav7399}.
\bibitem[{{Marini Bettolo Marconi} and Tarazona(1999)}]{MarconiT1999}
\bibinfo{author}{{Marini Bettolo Marconi}, U.}, \bibinfo{author}{Tarazona, P.},
  \bibinfo{year}{1999}.
\newblock \bibinfo{title}{Dynamic density functional theory of fluids}.
\newblock \bibinfo{journal}{Journal of Chemical Physics} \bibinfo{volume}{110},
  \bibinfo{pages}{8032--8044}.
\bibitem[{{Marini Bettolo Marconi} and Tarazona(2000)}]{MarconiT2000}
\bibinfo{author}{{Marini Bettolo Marconi}, U.}, \bibinfo{author}{Tarazona, P.},
  \bibinfo{year}{2000}.
\newblock \bibinfo{title}{Dynamic density functional theory of fluids}.
\newblock \bibinfo{journal}{Journal of Physics: Condensed Matter}
  \bibinfo{volume}{12}, \bibinfo{pages}{413--418}.
\bibitem[{Menon and Callender(2013)}]{MenonC2013}
\bibinfo{author}{Menon, T.}, \bibinfo{author}{Callender, C.},
  \bibinfo{year}{2013}.
\newblock \bibinfo{title}{Turn and face the strange … ch-ch-changes:
  Philosophical questions raised by phase transitions}, in:
  \bibinfo{editor}{Batterman, R.} (Ed.), \bibinfo{booktitle}{The {O}xford
  Handbook of Philosophy of Physics}. \bibinfo{publisher}{Oxford University
  Press}, \bibinfo{address}{Oxford}, pp. \bibinfo{pages}{190--223}.
\bibitem[{Menzel et~al.(2016)Menzel, Saha, Hoell and L{\"o}wen}]{MenzelSHL2016}
\bibinfo{author}{Menzel, A.M.}, \bibinfo{author}{Saha, A.},
  \bibinfo{author}{Hoell, C.}, \bibinfo{author}{L{\"o}wen, H.},
  \bibinfo{year}{2016}.
\newblock \bibinfo{title}{Dynamical density functional theory for
  microswimmers}.
\newblock \bibinfo{journal}{Journal of Chemical Physics} \bibinfo{volume}{144},
  \bibinfo{pages}{024115}.
\bibitem[{Mermin(1965)}]{Mermin1965}
\bibinfo{author}{Mermin, N.D.}, \bibinfo{year}{1965}.
\newblock \bibinfo{title}{Thermal properties of the inhomogeneous electron
  gas}.
\newblock \bibinfo{journal}{Physical Review} \bibinfo{volume}{137},
  \bibinfo{pages}{A1441--A1443}.
\bibitem[{Meyer et~al.(2019)Meyer, Voigtmann and Schilling}]{MeyerVS2019}
\bibinfo{author}{Meyer, H.}, \bibinfo{author}{Voigtmann, T.},
  \bibinfo{author}{Schilling, T.}, \bibinfo{year}{2019}.
\newblock \bibinfo{title}{On the dynamics of reaction coordinates in classical,
  time-dependent, many-body processes}.
\newblock \bibinfo{journal}{Journal of Chemical Physics} \bibinfo{volume}{150},
  \bibinfo{pages}{174118}.
\bibitem[{Mori(1965)}]{Mori1965}
\bibinfo{author}{Mori, H.}, \bibinfo{year}{1965}.
\newblock \bibinfo{title}{Transport, collective motion, and {B}rownian motion}.
\newblock \bibinfo{journal}{Progress of Theoretical Physics}
  \bibinfo{volume}{33}, \bibinfo{pages}{423--455}.
\bibitem[{Munakata(1989)}]{Munakata1989}
\bibinfo{author}{Munakata, T.}, \bibinfo{year}{1989}.
\newblock \bibinfo{title}{A dynamical extension of the density functional
  theory}.
\newblock \bibinfo{journal}{Journal of the Physical Society of Japan}
  \bibinfo{volume}{58}, \bibinfo{pages}{2434--2438}.
\bibitem[{Munakata(1994)}]{Munakata1994}
\bibinfo{author}{Munakata, T.}, \bibinfo{year}{1994}.
\newblock \bibinfo{title}{Time-dependent density-functional theory with {H}
  theorems}.
\newblock \bibinfo{journal}{Physical Review E} \bibinfo{volume}{50},
  \bibinfo{pages}{2347--2350}.
\bibitem[{Nakajima(1958)}]{Nakajima1958}
\bibinfo{author}{Nakajima, S.}, \bibinfo{year}{1958}.
\newblock \bibinfo{title}{On quantum theory of transport phenomena: steady
  diffusion}.
\newblock \bibinfo{journal}{Progress of Theoretical Physics}
  \bibinfo{volume}{20}, \bibinfo{pages}{948--959}.
\bibitem[{North(2011)}]{North2011}
\bibinfo{author}{North, J.}, \bibinfo{year}{2011}.
\newblock \bibinfo{title}{Time in thermodynamics}, in:
  \bibinfo{editor}{Callender, C.} (Ed.), \bibinfo{booktitle}{The Oxford
  Handbook of Philosophy of Time}. \bibinfo{publisher}{Oxford University
  Press}, pp. \bibinfo{pages}{312--350}.
\bibitem[{Penrose(1994)}]{Penrose1994}
\bibinfo{author}{Penrose, R.}, \bibinfo{year}{1994}.
\newblock \bibinfo{title}{On the second law of thermodynamics}.
\newblock \bibinfo{journal}{Journal of Statistical Physics}
  \bibinfo{volume}{77}, \bibinfo{pages}{217--221}.
\bibitem[{Price(1996)}]{Price1996}
\bibinfo{author}{Price, H.}, \bibinfo{year}{1996}.
\newblock \bibinfo{title}{Time's Arrow and Archimede's Point}.
\newblock \bibinfo{publisher}{Oxford University Press},
  \bibinfo{address}{Oxford}.
\bibitem[{{Rauscher} et~al.(2007){Rauscher}, {Dom{\'{\i}}nguez}, {Kr{\"u}ger}
  and {Penna}}]{RauscherDKP2007}
\bibinfo{author}{{Rauscher}, M.}, \bibinfo{author}{{Dom{\'{\i}}nguez}, A.},
  \bibinfo{author}{{Kr{\"u}ger}, M.}, \bibinfo{author}{{Penna}, F.},
  \bibinfo{year}{2007}.
\newblock \bibinfo{title}{A dynamic density functional theory for particles in
  a flowing solvent}.
\newblock \bibinfo{journal}{Journal of Chemical Physics} \bibinfo{volume}{127},
  \bibinfo{pages}{244906}.
\bibitem[{Rex and L{\"o}wen(2008)}]{RexL2008}
\bibinfo{author}{Rex, M.}, \bibinfo{author}{L{\"o}wen, H.},
  \bibinfo{year}{2008}.
\newblock \bibinfo{title}{Dynamical density functional theory with hydrodynamic
  interactions and colloids in unstable traps}.
\newblock \bibinfo{journal}{Physical Review Letters} \bibinfo{volume}{101},
  \bibinfo{pages}{148302}.
\bibitem[{Ridderbos and Redhead(1998)}]{RidderbosR1998}
\bibinfo{author}{Ridderbos, T.M.}, \bibinfo{author}{Redhead, M.L.G.},
  \bibinfo{year}{1998}.
\newblock \bibinfo{title}{The spin-echo experiments and the second law of
  thermodynamics}.
\newblock \bibinfo{journal}{Foundations of Physics} \bibinfo{volume}{28},
  \bibinfo{pages}{1237--1270}.
\bibitem[{Robbins et~al.(2011)Robbins, Archer and Thiele}]{RobbinsAT2011}
\bibinfo{author}{Robbins, M.J.}, \bibinfo{author}{Archer, A.J.},
  \bibinfo{author}{Thiele, U.}, \bibinfo{year}{2011}.
\newblock \bibinfo{title}{Modelling the evaporation of thin films of colloidal
  suspensions using dynamical density functional theory}.
\newblock \bibinfo{journal}{Journal of Physics: Condensed Matter}
  \bibinfo{volume}{23}, \bibinfo{pages}{415102}.
\bibitem[{Robertson(2020)}]{Robertson2018}
\bibinfo{author}{Robertson, K.}, \bibinfo{year}{2020}.
\newblock \bibinfo{title}{Asymmetry, abstraction, and autonomy: Justifying
  coarse-graining in statistical mechanics}.
\newblock \bibinfo{journal}{British Journal for the Philosophy of Science}
  \bibinfo{volume}{71}, \bibinfo{pages}{547--579}.
\bibitem[{Runge and Gross(1984)}]{RungeG1984}
\bibinfo{author}{Runge, E.}, \bibinfo{author}{Gross, E.K.U.},
  \bibinfo{year}{1984}.
\newblock \bibinfo{title}{Density-functional theory for time-dependent
  systems}.
\newblock \bibinfo{journal}{Physical Review Letters} \bibinfo{volume}{52},
  \bibinfo{pages}{997--1000}.
\bibitem[{Schmidt and Brader(2013)}]{SchmidtB2013}
\bibinfo{author}{Schmidt, M.}, \bibinfo{author}{Brader, J.M.},
  \bibinfo{year}{2013}.
\newblock \bibinfo{title}{Power functional theory for {B}rownian dynamics}.
\newblock \bibinfo{journal}{Journal of Chemical Physics} \bibinfo{volume}{138},
  \bibinfo{pages}{214101}.
\bibitem[{Sklar(1995)}]{Sklar1995}
\bibinfo{author}{Sklar, L.}, \bibinfo{year}{1995}.
\newblock \bibinfo{title}{Physics and Chance: Philosophical issues in the
  foundations of statistical mechanics}.
\newblock \bibinfo{publisher}{Cambridge University Press}.
\bibitem[{Sober(2020)}]{Sober2020}
\bibinfo{author}{Sober, E.}, \bibinfo{year}{2020}.
\newblock \bibinfo{title}{Histories, dynamical laws, and initial conditions-
  invariance under time-reversibility and its failure in {M}arkov processes,
  with application to the second law of thermodynamics and the past
  hypothesis}.
\newblock \bibinfo{journal}{Studies in History and Philosophy of Modern
  Physics} \bibinfo{volume}{69}, \bibinfo{pages}{26--31}.
\bibitem[{Streater(2009)}]{Streater2009}
\bibinfo{author}{Streater, R.}, \bibinfo{year}{2009}.
\newblock \bibinfo{title}{Statistical Dynamics: A Stochastic Approach to
  Nonequilibrium Thermodynamics}.
\newblock \bibinfo{publisher}{Imperial College Press},
  \bibinfo{address}{London}.
\bibitem[{{te Vrugt} et~al.(2020a){te Vrugt}, Bickmann and
  Wittkowski}]{teVrugtBW2020}
\bibinfo{author}{{te Vrugt}, M.}, \bibinfo{author}{Bickmann, J.},
  \bibinfo{author}{Wittkowski, R.}, \bibinfo{year}{2020}a.
\newblock \bibinfo{title}{Effects of social distancing and isolation on
  epidemic spreading modeled via dynamical density functional theory}.
\newblock \bibinfo{journal}{Nature Communications} \bibinfo{volume}{11},
  \bibinfo{pages}{5576}.
\bibitem[{{te Vrugt} et~al.(2020b){te Vrugt}, L{\"o}wen and
  Wittkowski}]{teVrugtLW2020}
\bibinfo{author}{{te Vrugt}, M.}, \bibinfo{author}{L{\"o}wen, H.},
  \bibinfo{author}{Wittkowski, R.}, \bibinfo{year}{2020}b.
\newblock \bibinfo{title}{Classical dynamical density functional theory: from
  fundamentals to applications}.
\newblock \bibinfo{journal}{Advances in Physics} \bibinfo{volume}{69},
  \bibinfo{pages}{121--247}.
\bibitem[{{te Vrugt} and Wittkowski(2019)}]{teVrugtW2019}
\bibinfo{author}{{te Vrugt}, M.}, \bibinfo{author}{Wittkowski, R.},
  \bibinfo{year}{2019}.
\newblock \bibinfo{title}{{Mori-Zwanzig projection operator formalism for
  far-from-equilibrium systems with time-dependent Hamiltonians}}.
\newblock \bibinfo{journal}{Physical Review E} \bibinfo{volume}{99},
  \bibinfo{pages}{062118}.
\bibitem[{{te Vrugt} and Wittkowski(2020a)}]{teVrugtW2019b}
\bibinfo{author}{{te Vrugt}, M.}, \bibinfo{author}{Wittkowski, R.},
  \bibinfo{year}{2020}a.
\newblock \bibinfo{title}{Projection operators in statistical mechanics: a
  pedagogical approach}.
\newblock \bibinfo{journal}{European Journal of Physics} \bibinfo{volume}{41},
  \bibinfo{pages}{045101}.
\bibitem[{{te Vrugt} and Wittkowski(2020b)}]{teVrugtW2020}
\bibinfo{author}{{te Vrugt}, M.}, \bibinfo{author}{Wittkowski, R.},
  \bibinfo{year}{2020}b.
\newblock \bibinfo{title}{Relations between angular and {C}artesian
  orientational expansions}.
\newblock \bibinfo{journal}{AIP Advances} \bibinfo{volume}{10},
  \bibinfo{pages}{035106}.
\bibitem[{Thiele et~al.(2019)Thiele, Frohoff-H{\"u}lsmann, Engelnkemper,
  Knobloch and Archer}]{ThieleFHEKA2019}
\bibinfo{author}{Thiele, U.}, \bibinfo{author}{Frohoff-H{\"u}lsmann, T.},
  \bibinfo{author}{Engelnkemper, S.}, \bibinfo{author}{Knobloch, E.},
  \bibinfo{author}{Archer, A.J.}, \bibinfo{year}{2019}.
\newblock \bibinfo{title}{First order phase transitions and the thermodynamic
  limit}.
\newblock \bibinfo{journal}{New Journal of Physics} \bibinfo{volume}{21},
  \bibinfo{pages}{123021}.
\bibitem[{Treffenst{\"a}dt and Schmidt(2020)}]{TreffenstadtS2020}
\bibinfo{author}{Treffenst{\"a}dt, L.L.}, \bibinfo{author}{Schmidt, M.},
  \bibinfo{year}{2020}.
\newblock \bibinfo{title}{Memory-induced motion reversal in {B}rownian
  liquids}.
\newblock \bibinfo{journal}{Soft Matter} \bibinfo{volume}{16},
  \bibinfo{pages}{1518--1526}.
\bibitem[{Uffink(2001)}]{Uffink2001}
\bibinfo{author}{Uffink, J.}, \bibinfo{year}{2001}.
\newblock \bibinfo{title}{Bluff your way in the second law of thermodynamics}.
\newblock \bibinfo{journal}{Studies in History and Philosophy of Modern
  Physics} \bibinfo{volume}{32}, \bibinfo{pages}{305--394}.
\bibitem[{Uffink(2007)}]{Uffink2006}
\bibinfo{author}{Uffink, J.}, \bibinfo{year}{2007}.
\newblock \bibinfo{title}{Compendium of the foundations of classical
  statistical physics}, in: \bibinfo{editor}{Butterfield, J.},
  \bibinfo{editor}{Earman, J.} (Eds.), \bibinfo{booktitle}{Philosophy of
  Physics}. \bibinfo{publisher}{Elsevier}, \bibinfo{address}{Netherlands}, pp.
  \bibinfo{pages}{923--1074}.
\bibitem[{Wallace(2011)}]{Wallace2011}
\bibinfo{author}{Wallace, D.}, \bibinfo{year}{2011}.
\newblock \bibinfo{title}{The logic of the past hypothesis}.
\newblock \bibinfo{journal}{preprint, available at
  http://philsci-archive.pitt.edu/8894/} .
\bibitem[{Wallace(2013)}]{Wallace2013}
\bibinfo{author}{Wallace, D.}, \bibinfo{year}{2013}.
\newblock \bibinfo{title}{The arrow of time in physics}, in:
  \bibinfo{editor}{Dyke, H.}, \bibinfo{editor}{Bardon, A.} (Eds.),
  \bibinfo{booktitle}{A Companion to the Philosophy of Time}.
  \bibinfo{publisher}{John Wiley \& Sons}. chapter~\bibinfo{chapter}{16}, pp.
  \bibinfo{pages}{262--281}.
\bibitem[{Wallace(2015)}]{Wallace2015}
\bibinfo{author}{Wallace, D.}, \bibinfo{year}{2015}.
\newblock \bibinfo{title}{The quantitative content of statistical mechanics}.
\newblock \bibinfo{journal}{Studies in History and Philosophy of Modern
  Physics} \bibinfo{volume}{52}, \bibinfo{pages}{285--293}.
\bibitem[{Wensink and L{\"o}wen(2008)}]{WensinkL2008}
\bibinfo{author}{Wensink, H.H.}, \bibinfo{author}{L{\"o}wen, H.},
  \bibinfo{year}{2008}.
\newblock \bibinfo{title}{Aggregation of self-propelled colloidal rods near
  confining walls}.
\newblock \bibinfo{journal}{Physical Review E} \bibinfo{volume}{78},
  \bibinfo{pages}{031409}.
\bibitem[{Wittkowski and L{\"o}wen(2011)}]{WittkowskiL2011}
\bibinfo{author}{Wittkowski, R.}, \bibinfo{author}{L{\"o}wen, H.},
  \bibinfo{year}{2011}.
\newblock \bibinfo{title}{Dynamical density functional theory for colloidal
  particles with arbitrary shape}.
\newblock \bibinfo{journal}{Molecular Physics} \bibinfo{volume}{109},
  \bibinfo{pages}{2935--2943}.
\bibitem[{Wittkowski et~al.(2012)Wittkowski, L{\"o}wen and
  Brand}]{WittkowskiLB2012}
\bibinfo{author}{Wittkowski, R.}, \bibinfo{author}{L{\"o}wen, H.},
  \bibinfo{author}{Brand, H.R.}, \bibinfo{year}{2012}.
\newblock \bibinfo{title}{Extended dynamical density functional theory for
  colloidal mixtures with temperature gradients}.
\newblock \bibinfo{journal}{Journal of Chemical Physics} \bibinfo{volume}{137},
  \bibinfo{pages}{224904}.
\bibitem[{Wittkowski et~al.(2013)Wittkowski, L{\"o}wen and
  Brand}]{WittkowskiLB2013}
\bibinfo{author}{Wittkowski, R.}, \bibinfo{author}{L{\"o}wen, H.},
  \bibinfo{author}{Brand, H.R.}, \bibinfo{year}{2013}.
\newblock \bibinfo{title}{Microscopic approach to entropy production}.
\newblock \bibinfo{journal}{Journal of Physics A: Mathematical and Theoretical}
  \bibinfo{volume}{46}, \bibinfo{pages}{355003}.
\bibitem[{Wittmann and Brader(2016)}]{WittmannB2016}
\bibinfo{author}{Wittmann, R.}, \bibinfo{author}{Brader, J.M.},
  \bibinfo{year}{2016}.
\newblock \bibinfo{title}{Active {B}rownian particles at interfaces: an
  effective equilibrium approach}.
\newblock \bibinfo{journal}{Europhysics Letters} \bibinfo{volume}{114},
  \bibinfo{pages}{68004}.
\bibitem[{Wittmann et~al.(2021)Wittmann, L{\"o}wen and Brader}]{WittmannLB2020}
\bibinfo{author}{Wittmann, R.}, \bibinfo{author}{L{\"o}wen, H.},
  \bibinfo{author}{Brader, J.M.}, \bibinfo{year}{2021}.
\newblock \bibinfo{title}{Order-preserving dynamics in one dimension –
  single-file diffusion and caging from the perspective of dynamical density
  functional theory}.
\newblock \bibinfo{journal}{Molecular Physics} ,
  \bibinfo{pages}{e1867250}\bibinfo{note}{\quad
  doi:10.1080/00268976.2020.1867250}.
\bibitem[{Yoshimori(2005)}]{Yoshimori2005}
\bibinfo{author}{Yoshimori, A.}, \bibinfo{year}{2005}.
\newblock \bibinfo{title}{Microscopic derivation of time-dependent density
  functional methods}.
\newblock \bibinfo{journal}{Physical Review E} \bibinfo{volume}{71},
  \bibinfo{pages}{031203}.
\bibitem[{Zeh(2007)}]{Zeh1989}
\bibinfo{author}{Zeh, H.D.}, \bibinfo{year}{2007}.
\newblock \bibinfo{title}{The physical basis of the direction of time (5th
  edition)}.
\newblock \bibinfo{publisher}{Springer Verlag}, \bibinfo{address}{Berlin
  Heidelberg}.
\bibitem[{Zwanzig(1960)}]{Zwanzig1960}
\bibinfo{author}{Zwanzig, R.}, \bibinfo{year}{1960}.
\newblock \bibinfo{title}{Ensemble method in the theory of irreversibility}.
\newblock \bibinfo{journal}{Journal of Chemical Physics} \bibinfo{volume}{33},
  \bibinfo{pages}{1338--1341}.

\end{thebibliography}

\end{document}